\documentclass[aps,jmp,reprint]{revtex4-1}

\usepackage{graphicx}
\usepackage{dcolumn}% Align table columns on decimal point
\usepackage{bm}
\usepackage{amsmath}
\usepackage{amssymb}
\usepackage{amsbsy}
\usepackage{amsfonts}
\usepackage{mathtools}
\newcommand{\state}{Statement}
\newtheorem{theorem}{\state}

\begin{document}

%\preprint{}

\title{On the sub-shock formation in extended thermodynamics}


\author{Shigeru Taniguchi$^1$ and Tommaso Ruggeri$^2$}
\email[]{taniguchi.shigeru@kct.ac.jp, tommaso.ruggeri@unibo.it}
%\thanks{}
\affiliation{
$^1$Department of Creative Engineering, National Institute of Technology, Kitakyushu College, Japan\\
$^2$Department of Mathematics $\&$ Alma Mater Research Center on Applied Mathematics,  \\ University of Bologna, Bologna, Italy
}

              \date{\today}

\begin{abstract}
In hyperbolic dissipative systems, the solution of the shock structure is not always continuous and a discontinuous part (sub-shock) appears when the velocity of the shock wave is greater than a critical value.
In principle, the sub-shock may occur when the shock velocity $s$ reaches one of  the characteristic eigenvalues of the hyperbolic system. 
Nevertheless, Rational Extended Thermodynamics (ET) for a rarefied monatomic gas predicts the sub-shock formation only when $s$ exceeds the maximum characteristic velocity of the system evaluated in the unperturbed state $\lambda^{\max}_0$. 
This fact agrees with a general theorem asserting that continuous shock structure cannot exist for $s >\lambda^{\max}_0 $. 
In the present paper, first, the shock structure is numerically analyzed on the basis of ET for a rarefied polyatomic gas with $14$ independent fields. 
It is shown that, also in this case, the shock structure is still continuous when $s$ meets characteristic velocities except for the maximum one and therefore the sub-shock appears only when $s >\lambda^{\max}_0 $.  
This example reinforces the conjecture that, the differential systems of ET theories have the special characteristics such that the sub-shock appears only for $s$ greater than the unperturbed maximum characteristic velocity.
However, in the second part of the  paper, we construct a counterexample of this conjecture by using a simple $2 \times 2$ hyperbolic dissipative system which satisfies all requirements of ET. 
In contrast to previous results, we show  the clear sub-shock formation with a slower shock velocity than the maximum unperturbed characteristic velocity. 
\begin{description}
\item[PACS numbers] 47.40.-x, 05.70.Ln, 47.45.-n
\end{description}
\end{abstract}

%\pacs{05.70.Ln}
%\keywords{}
%\showpacs
\maketitle

\section{Introduction}

Hyperbolic dissipative systems, which are sometimes called as hyperbolic systems with relaxation in the mathematical community, describe a large class of the physical systems and appear in many fields, in particular, in the field of non-equilibrium thermodynamics within the framework of so-called Rational Extended Thermodynamics (hereafter, for simplicity, referred to as ET, instead of RET) ~\cite{MullerRuggeri,RuggeriSugiyama}. 
In (parabolic or hyperbolic) dissipative systems, the shock wave is represented by a solution of the type of  traveling waves that is called  \emph{shock structure} because it predicts a  thickness of the shock wave.  
In contrast to the parabolic system with the Navier-Stokes  and  Fourier (NSF) constitutive equations obtained in the framework of Thermodynamics of Irreversible Processes (TIP),  the hyperbolic dissipative system predicts, in general, the formation of a \emph{sub-shock}. 
In  other words, the shock structure is not always continuous and a discontinuous part (sub-shock) appears when the velocity of the shock wave $s$ is greater than a critical value. 

For the shock structure in rarefied monatomic gases, the following features have been reported in literature: 
Grad proposed the moment method of closure of the field equations~\cite{Grad1949} and showed that the discontinuity (sub-shock) may appear in the so-called Grad-13 moment system when the Mach number is greater than 1.65, which corresponds to the value of $s$ reaching the maximum characteristic velocity evaluated in equilibrium unperturbed state ~\cite{Grad1952}.  
Ruggeri showed that, for any hyperbolic system of  balance laws, the shock structure  becomes in principle singular when the shock velocity $s$ meets a characteristic velocity and therefore the sub-shock seems to appear when $s$ meets all the supersonic characteristic velocities of the hyperbolic system~\cite{Ruggeri1993}. 

In order to check the theoretical prediction of the sub-shock formation, Weiss performed numerical calculations of the shock structure in a rarefied monatomic gas on the basis of ET with 13, 14 and 21 independent variables with the use of the assumption of the Maxwellian molecule for production terms. 
The numerical results showed that, except for the maximum characteristic velocity, the singular points become regular and continuous solution is obtained until $s$ reaches the maximum characteristic velocity. 
Weiss concluded, as a conjecture, that for any number of moments  the sub-shock appears only after the maximum characteristic velocity, at least numerically~\cite{Weiss}. 
This conjecture was reinforced by a theorem of Boillat and Ruggeri in which it was proven that, for hyperbolic system of balance laws satisfying the convexity of the entropy, no continuous solution exists with larger shock velocity $s$ than the maximum characteristic velocity evaluated in the unperturbed state $\lambda^{\max}_0$ ~\cite{Breakdown}. 

However, there is no  mathematical proof about the absence of the sub-shock when the shock velocity is slower than the maximum characteristic velocity.
There still remain the following questions: ``Is the above conjecture valid for all systems satisfying the requirements of ET theory?'' and 
``Are there any possibilities to have the sub-shock with slower characteristic velocity than the maximum characteristic velocity?''
These questions are interesting not only mathematically but also physically due to the following recent progresses:  

(a) Extended thermodynamics of polyatomic gases has been developed~\cite{ET14,ET6,NLET6}. 
The ET theory with 14 independent variables (ET$_{14}$) explains the shock structure in rarefied polyatomic gases where the internal modes, namely, rotational and vibrational modes, are partially excited~\cite{ET14shock}.
In particular, ET$_{14}$ can explain the structure composed of thin and thick layers~\cite{VincentiKruger,Zeldovich} in a fully consistent way~\cite{ET14shock} in contrast to previous Bethe-Teller theory~\cite{BetheTeller}.  
It is also shown that the very steep change in the thin layer may be described as a sub-shock within the resolution of the simplified ET theory with only $6$ independent fields (ET$_6$)~\cite{ET14shock,ET6shock,NLET6shock}. 
The numerical results based on the kinetic theory also support the theoretical predictions by the ET theories quantitatively~\cite{Kosuge}. 
Therefore the sub-shock formation does not necessarily imply the violation of the validity range of the ET theory in a polyatomic gas and the sub-shock may have the physical meaning in this kind of problems. 

(b)  In the context of  a binary mixture of Eulerian monatomic gases, the sub-shock formation with slower shock velocity than the maximum  unperturbed characteristic velocity and the multiple sub-shock was observed via numerical analysis~\cite{Bisi1,FMR}. 
However, the system of balance equations for binary mixtures is very special  because the field equations for each component have exactly the same form of a single fluid and  the coupling effect is only through the production terms that take the mechanical and thermal diffusions into account.

In the present paper, in order to understand the problematics more deeply, we first reconsider the shock structure in a rarefied polyatomic gas predicted by ET$_{14}$ and it will be shown that, also in this case, the singular points where $s$ reaches  slower characteristic velocities may become regular
%, as in the case of a monatomic gas, 
and the sub-shock appears only when the shock velocity is greater than the maximum characteristic velocity in the unperturbed state. 
This example reinforces the conjecture that,  the differential systems of  ET theories have the special characteristics such that the sub-shock occurs only for $s$ greater than the unperturbed maximum characteristic velocity.

However, in the  second part of the paper, we construct a counterexample of this conjecture by using a simple $2 \times 2$ hyperbolic dissipative system that satisfies all requirements of extended thermodynamics, that is, the entropy inequality, concavity of the entropy, sub-characteristic condition and Shizuta-Kawashima condition.
In contrast to previous results, we show clearly the sub-shock formation with a shock velocity slower than the maximum characteristic velocity. 
Moreover,  multiple sub-shock is also observed in this simple system. 

Final section is devoted to the concluding remarks and the discussion on some open problems. 

\section{Shock-structure problem}
The system of field equations of ET in one space dimension belongs to a particular case of general first order hyperbolic quasi-linear system of balance laws: 
\begin{equation}\label{sistemagenerale}
\frac{\partial \mathbf{U}}{\partial t} + \frac{\partial \mathbf{F}(\mathbf{U})}{\partial x} = \mathbf{f}(\mathbf{U}), 
\end{equation}
where $\mathbf{U}$, $\mathbf{F}$ and $\mathbf{f}$ are column vectors of $R^N$. 
Here $\mathbf{U}(x,t)$ is the unknown field vector with $x$ and $t$ being, respectively, the  space and time.

Let us consider a solution of \eqref{sistemagenerale} representing a shock structure, that is, the field variable $\mathbf{U}$ depends only on a single variable $z$ (traveling wave): 
\begin{equation*}
\mathbf{U}\equiv \mathbf{U}(z), \qquad z= x - s t
\end{equation*}
with constant equilibrium  boundary conditions at  infinity:
\begin{equation}\label{contorno}
 \lim_{z\rightarrow  + \infty  } \mathbf{U} =\mathbf{U}_0, \qquad \lim_{z \rightarrow  - \infty } \mathbf{U} = \mathbf{U}_1, 
\end{equation} 
where
\begin{equation*}
 \mathbf{f}(\mathbf{U}_0) = \mathbf{f}(\mathbf{U}_1)= 0.
\end{equation*}
We call the state $\mathbf{U}_0$ as the {\it unperturbed state} and the state $\mathbf{U}_1$ as the {\it perturbed state}, respectively. 
Hereafter, the quantities with the subscript 0 represent the quantities evaluated in the unperturbed state and the quantities with subscript 1 represent the ones evaluated in the perturbed state. 
From \eqref{sistemagenerale}, we have the following ODE system:
\begin{equation}\label{struttura}
\left( \mathbf{A}( \mathbf{U}) - s \mathbf{I}
\right)\frac{d\mathbf{U}}{dz}= \mathbf{f}( \mathbf{U}), \qquad \mathbf{A}= \frac{\partial \mathbf{F}}{\partial \mathbf{U}}
\end{equation}
with boundary conditions given by \eqref{contorno}.

Following \cite{Breakdown}, by taking the typical features of extended thermodynamics into account, we may split the system \eqref{sistemagenerale} into the blocks of $M$  conservation laws and of $N-M$ balance equations as follows:   
\begin{equation}\label{blocks}
\begin{split}
&\frac{\partial \mathbf{V}(\mathbf{U})}{\partial t} + \frac{\partial \mathbf{P}(\mathbf{U})}{\partial x} = 0, \\
&\frac{\partial \mathbf{W}(\mathbf{U})}{\partial t} + \frac{\partial \mathbf{R}(\mathbf{U})}{\partial x} = \mathbf{g}(\mathbf{U}). 
\end{split}
\end{equation}
We may also choose the field variable $\mathbf{U}$ to coincide with the \emph{main field} by which the original system becomes  symmetric hyperbolic \cite{Boi,RS}: 
\begin{equation}
\mathbf{U} \equiv (\mathbf{v}, \mathbf{w})^T,
\end{equation}
where $\mathbf{v} \in R^M$ and $\mathbf{w} \in R^{N-M}$, such that \cite{SubSystem, Breakdown}:
\begin{equation}\label{equilibriow}
\mathbf{g} (\mathbf{v}, \mathbf{w}) = 0 \quad \Longleftrightarrow  \quad \mathbf{w}=0.
\end{equation}
The state with $\mathbf{w}=0$ represents the equilibrium state and we associate the system \eqref{blocks} with the corresponding  \emph{equilibrium subsystem}~\cite{SubSystem}:
\begin{equation}\label{eq:subgeneral}
\frac{\partial \mathbf{V}(\mathbf{v}, 0)}{\partial t} + \frac{\partial \mathbf{P}(\mathbf{v}, 0)}{\partial x} = 0. 
\end{equation}
Taking \eqref{blocks} into account, we may rewrite \eqref{struttura} as 
\begin{align}\label{eq:shockstructure_general}
\begin{split}
& \frac{d}{dz}\left\{- s \mathbf{V}(\mathbf{v}, \mathbf{w}) +\mathbf{P}(\mathbf{v}, \mathbf{w}) \right\} = 0, 
 \\
& - s \frac{d \mathbf{W}(\mathbf{v}, \mathbf{w})}{d z} + \frac{d \mathbf{R}(\mathbf{v}, \mathbf{w})}{d z} = \mathbf{g}(\mathbf{v}, \mathbf{w}). 
\end{split}
\end{align}
By integrating \eqref{eq:shockstructure_general}$_1$, we have
\begin{equation}\label{RHEq}
- s \mathbf{V}(\mathbf{v}, \mathbf{w}) +\mathbf{P}(\mathbf{v}, \mathbf{w}) = \text{const.} 
\end{equation}
and by taking the fact that unperturbed and perturbed states are constant states (see \eqref{contorno}), from \eqref{eq:shockstructure_general}$_2$ and \eqref{equilibriow}, we have 
\begin{equation}\label{wiszero}
\mathbf{w}_1 = \mathbf{w}_0 = 0
\end{equation}
and,  from \eqref{RHEq},
\begin{equation}\label{eq:RH_eqsub2}
- s \mathbf{V}(\mathbf{v}_0, 0) +\mathbf{P}(\mathbf{v}_0, 0) 
= - s \mathbf{V}(\mathbf{v}_1, 0) +\mathbf{P}(\mathbf{v}_1, 0). 
\end{equation}
This is nothing else the Rankine-Hugoniot (RH) conditions associated with the equilibrium subsystem \eqref{eq:subgeneral} and permits us to obtain $\mathbf{v}_1 \equiv \mathbf{v}_1(\mathbf{v}_0,s) $.
Therefore, once the unperturbed equilibrium state $\mathbf{v}_0$ and the shock velocity $s$ are given, the shock structure is obtained as the  solution of \eqref{RHEq} and \eqref{eq:shockstructure_general}$_2$ under the boundary conditions \eqref{wiszero} and \eqref{eq:RH_eqsub2}. 

According with \cite{Breakdown}, a singularity (sub-shock) may appear when a characteristic velocity $\lambda$,
which is the eigenvalue of the matrix  $\mathbf{A}$, meets the shock velocity (see \eqref{struttura}) for some $z$.
More precisely, let $\mathbf{U}_s({z})$ be a solution of \eqref{struttura} for a given $s$,
\begin{equation} \label{cnec}
\exists \, \bar{z }, \quad \text{such that}  \quad \lambda(\mathbf{U}_s(\bar{z})) =s.
\end{equation}
Assuming that, for a prescribed $s$, the solutions of \eqref{struttura} satisfy the following condition for any genuine non-linear eigenvalues $\lambda$:
\begin{equation} \label{inmezzo}
\lambda_0 \leq \lambda(\mathbf{U}_s({z})) \leq \lambda_1(s), \qquad \forall z  \in [-\infty,\infty].
\end{equation}
Then the necessary condition for a sub-shock with the shock velocity slower than $\lambda^{\max}_0 $,
is that, for some $s$, there exists an eigenvalue $\lambda$ such that
\begin{equation}\label{necessaria}
\lambda_0 < s <\lambda_1(s) <\lambda^{\max}_0.
\end{equation}
In fact, if \eqref{necessaria} is true, from \eqref{inmezzo},  \eqref{cnec} holds for continuity reason.

We notice that, if we increase the shock velocity more, such that $s > \lambda^{\max}_0$, the sub-shock corresponding to the fastest mode also becomes admissible and therefore we may expect that there exist two or more sub-shocks.

\section{Sub-shock formation in a rarefied polyatomic gas}
Let us analyze the shock structure in a rarefied polyatomic gas based on extended thermodynamics with 14 fields  (ET$_{14}$); the mass density $\rho$, the velocity $v_i$, the temperature $T$, the dynamic (non-equilibrium) pressure $\Pi$, the shear stress $\sigma_{\langle ij \rangle}$ and the heat flux $q_i$, where $i, j = 1,2,3$ and the angular brackets in $\sigma_{\langle ij \rangle}$ indicate that the shear stress is symmetric traceless tensor.
The ET$_{14}$ theory is the the simplest and natural extension of the Navier-Stokes and Fourier (NSF) theory and  ET$_{14}$ includes NSF as a special case. 

We adopt the caloric and thermal equations of state for a rarefied polyatomic gas.  
The specific internal energy $\varepsilon$ and the (equilibrium) pressure $p$ are expressed by
\begin{equation*}
\varepsilon = \frac{D}{2}\frac{k_B}{m}T, \quad p = \frac{k_B}{m} \rho T, 
\end{equation*}
where $D$, $k_B$ and $m$ are, respectively, the degrees of freedom of a molecule, the Boltzmann constant and the mass of a molecule. 
Hereafter, we consider a polytropic gas, that is, the specific heat is assumed to be constant ($D$ is constant). 
For the case of a non-polytropic rarefied gas, the shock structure was studied in \cite{ET14shock}. 

We focus on the one-dimensional (plane) shock waves propagating along the $x$-axis where the vectorial and tensorial quantities are given by
 \begin{align*}
  \label{1dim}
 \begin{split}
 &v_i \equiv \left(
 \begin{array}{c}
  v\\ 0\\ 0
   \end{array}
 \right), 
   \sigma_{\langle ij\rangle} \equiv \left(
 \begin{array}{ccc}
  \sigma & 0 &0\\
   0& -\frac{1}{2}\sigma & 0\\
  0& 0 & -\frac{1}{2}\sigma
   \end{array}
 \right), 
 q_i \equiv \left(
 \begin{array}{c}
  q\\ 0\\ 0
   \end{array}
 \right)
  \end{split}
 \end{align*}
and in this case, the independent variables are $\mathbf{U} \equiv (\rho, v, T, \Pi, \sigma, q)^T$. 
\begin{figure*}[]
	\begin{center}
		\includegraphics[width=0.4\hsize] {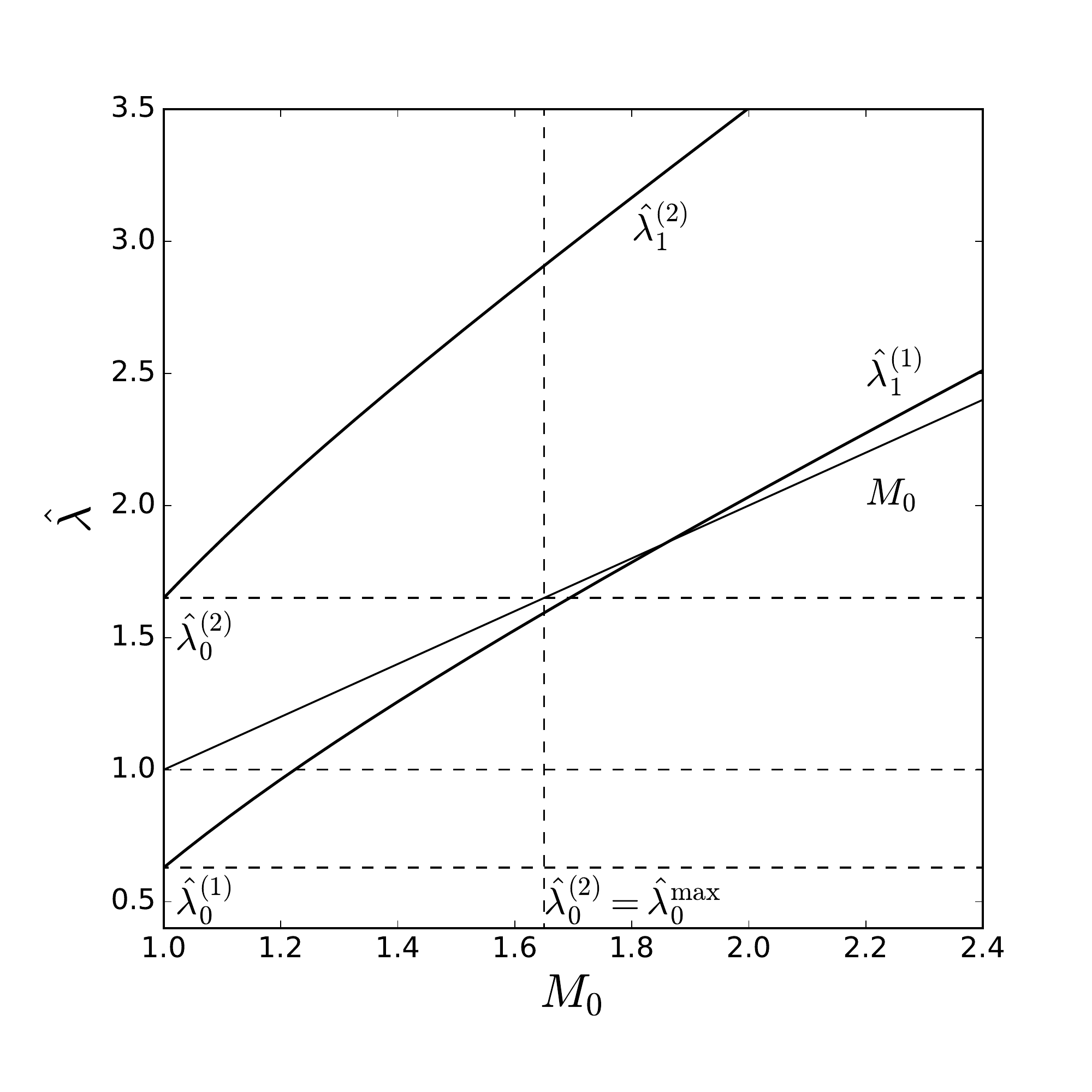}
		\includegraphics[width=0.4\hsize] {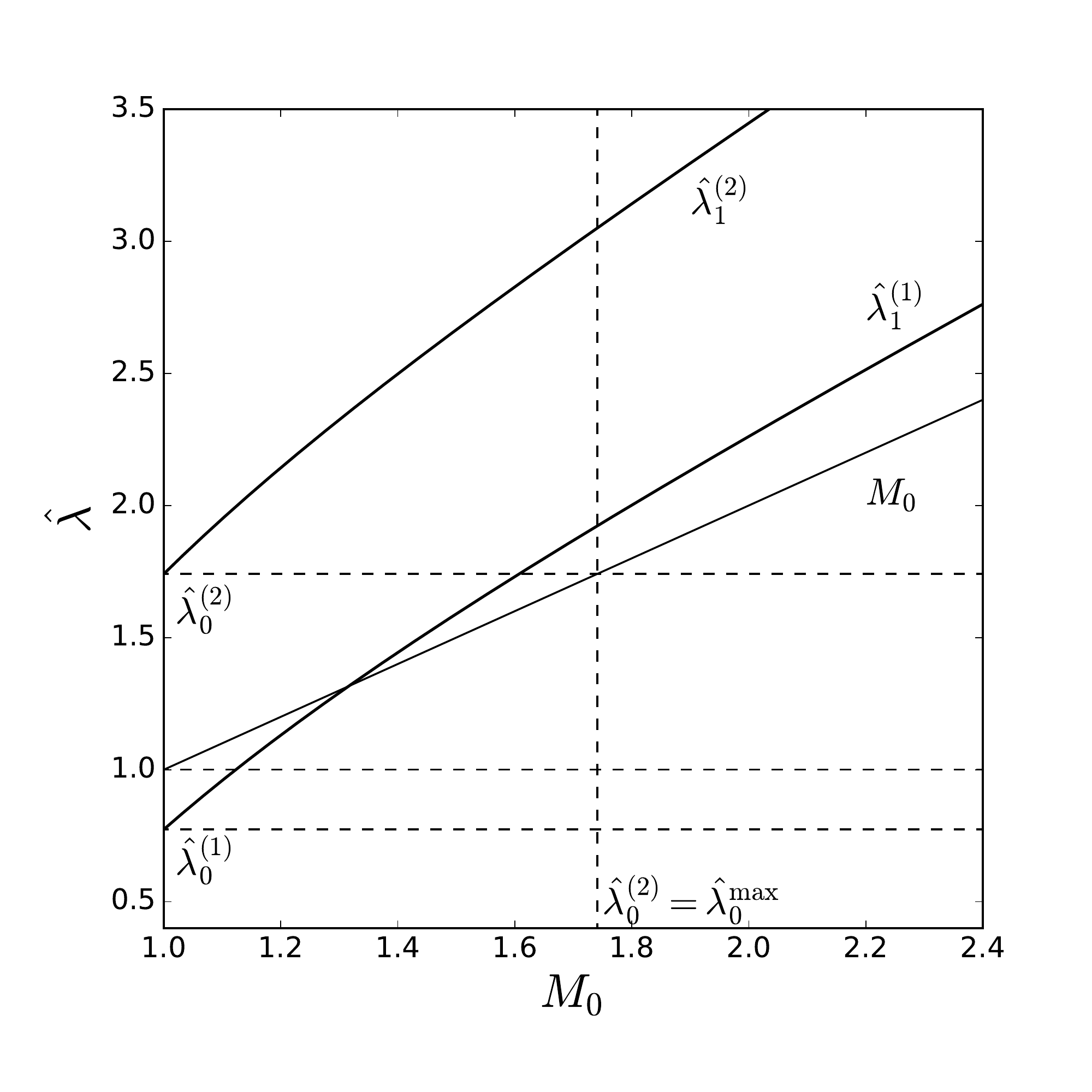}
		\caption{
			Mach number dependences of the dimensionless characteristic velocities in the perturbed state for $D=3$ (left) and for $D=7$ (right). 
		}
		\label{fig:ET14_diagram_D3}
	\end{center}
\end{figure*}
The field equations of  ET$_{14}$  are summarized as follows:~\cite{ET14}
\begin{equation}\label{eq:ET14}
\begin{split}
&\frac{\partial \rho}{\partial t}+\frac{\partial}{\partial  x}( \rho v ) = 0,\\
&\frac{\partial \rho v}{\partial t} + \frac{\partial }{\partial x}( p + \Pi - \sigma + \rho v^2 ) = 0,\\
&\frac{\partial}{\partial t} ( 2\rho \varepsilon +\rho v^2 ) + \\
& + \frac{\partial}{\partial x}\left\{2\rho \varepsilon v+ 2(p+\Pi-\sigma)v  + \rho v^3 + 2q\right\} = 0,\\
&\frac{\partial}{\partial t}\left\{ 3(p+\Pi) +\rho v^2 \right\} + \\
& + \frac{\partial}{\partial x}\left\{ (5p+5\Pi-2\sigma)v + \rho v^3 + \frac{5}{1+\hat{c}_v}q \right\} 
= -\frac{3\Pi}{\tau_{\Pi}}, \\
&\frac{\partial}{\partial t}( p+\Pi-\sigma +\rho v^2 ) + \\
& + \frac{\partial}{\partial x}\left\{ 3(p+\Pi-\sigma)v + \rho v^3 + \frac{3}{1+\hat{c}_v}q \right\} 
= \frac{\sigma}{\tau_{S}} -\frac{\Pi}{\tau_{\Pi}}, \\
&\frac{\partial}{\partial t}\left\{ 2\rho \varepsilon v+ 2(p+\Pi-\sigma)v  + \rho v^3 + 2q \right\} + \\
& + \frac{\partial}{\partial x}\bigg\{ 2\rho \varepsilon v^2+5(p+\Pi-\sigma)v^2 + \rho v^4 + \\
&\qquad + 2 \left(  \varepsilon + \frac{k_B}{m}T\right)p + 2 \left(  \varepsilon + 2\frac{k_B}{m}T\right)(\Pi - \sigma) + \\
&\qquad + \frac{10+4\hat{c}_v}{1+\hat{c}_v}q v \bigg\}\\
& = -2 \left\{\frac{q}{\tau_q}+\left(\frac{\Pi}{\tau_{\Pi}} - \frac{\sigma}{\tau_{S}}\right)v\right\}, 
\end{split}
\end{equation}
\begin{figure*}[]
	\begin{center}
		\includegraphics[width=0.49\hsize] {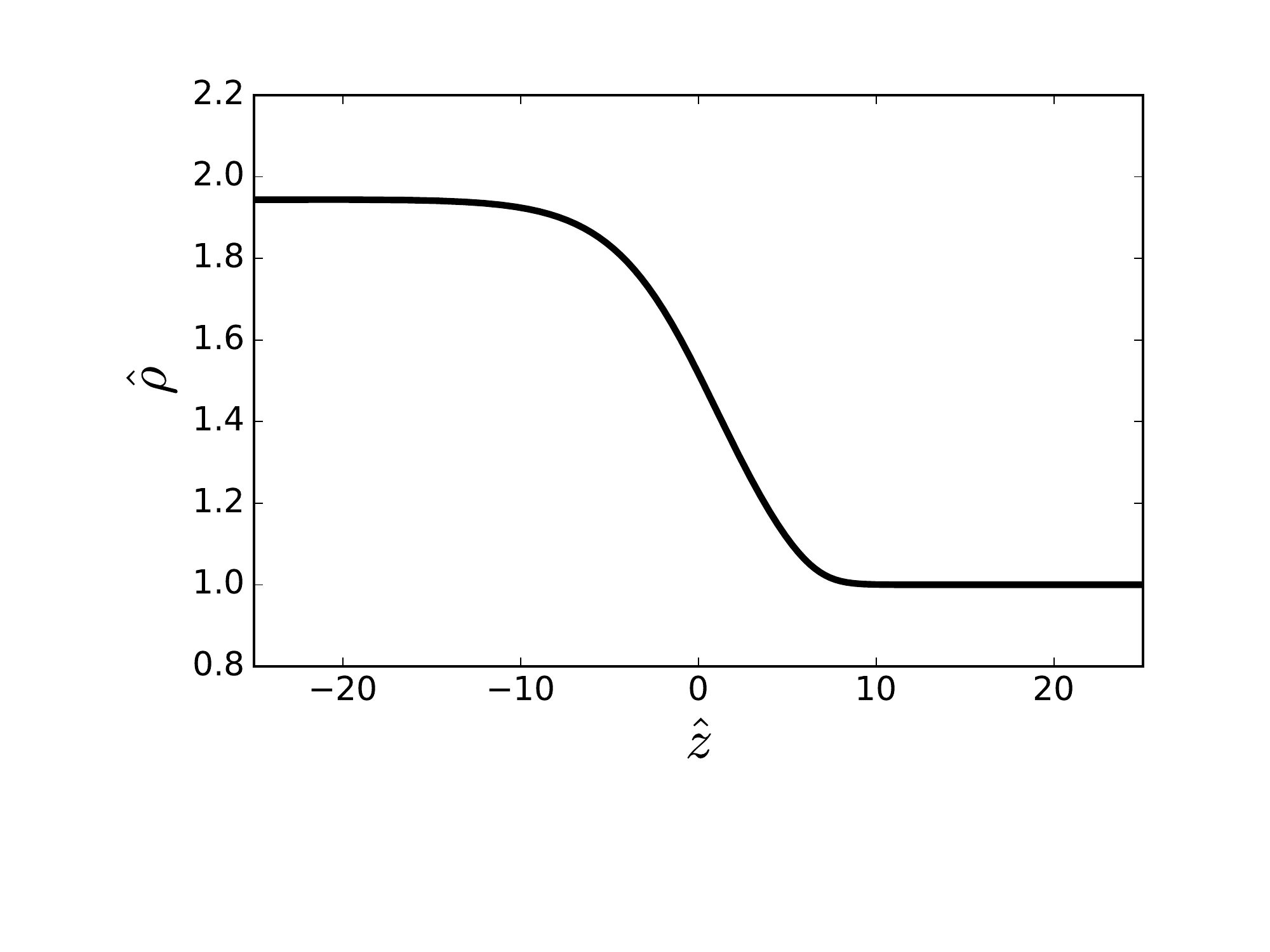}
		\includegraphics[width=0.49\hsize] {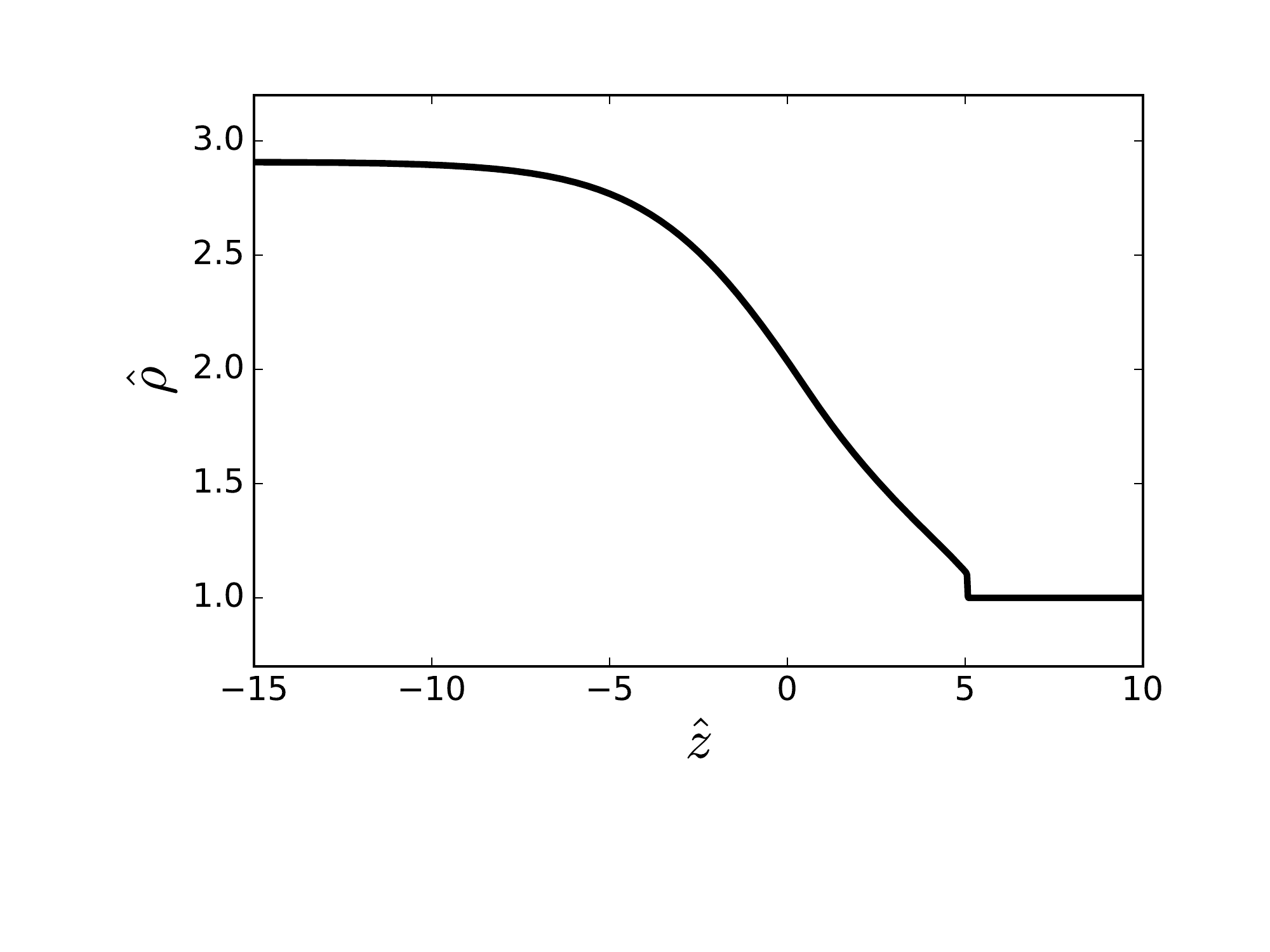}
		\caption{
			Profiles of the dimensionless mass density $\hat{\rho} \equiv \rho/\rho_0$ with $\rho_0$ being the mass density in the unperturbed state. 
			Here $\hat{z}$ is the dimensionless position defined by $\hat{z} \equiv z/(c_0 \tau_{\Pi})$
			and $D=7$. 
			$M_0=1.5$ (left) and $M_0=2$ (right). 
		}
		\label{fig:rho_ET14}
	\end{center}
\end{figure*}
where $\tau_{\Pi}$, $\tau_{S}$, and $\tau_{q}$ are the relaxation times for the dynamic pressure, 
the shear stress, and the heat flux, respectively. 
Here $\hat{c}_v$ is the dimensionless specific heat defined by $\hat{c_v}\equiv (m/k_B) c_v$ with $c_v$ being the specific heat and in the present case $\hat{c}_v=D/2$. 
The equilibrium state of \eqref{eq:ET14} is achieved when $\Pi=\sigma=q=0$.
The characteristic velocities in the equilibrium state $\lambda_E$  are \cite{ET14linear,RuggeriSugiyama}:
\begin{equation}\label{lame}
\frac{\lambda_E-v}{c} 
= 0, 0, \pm \Delta^{(1)}  , \pm \Delta^{(2)} , 
\end{equation}
where 
\begin{align*}\label{Deltina}
\begin{split}
&
\Delta^{(1)}   = \sqrt{\frac{\hat{c}_v\left( 7 + 4 \hat{c}_v - \sqrt{37 + 32 \hat{c}_v + 4 \hat{c}_v^2}\right)}
	{2 ( 1+\hat{c}_v )^2}}, \\
& \Delta^{(2)}  = \sqrt{\frac{\hat{c}_v\left( 7 + 4 \hat{c}_v + \sqrt{37 + 32 \hat{c}_v + 4 \hat{c}_v^2}\right)}
	{2 ( 1+\hat{c}_v )^2}}, 
\end{split}
\end{align*}
and  $c$ is the sound velocity:
\begin{equation}\label{soundv}
c = \sqrt{\gamma\frac{k_B}{m}T}. 
\end{equation}
Here $\gamma$ is the ratio of specific heats related with $\hat{c}_v$ and $D$  by the following relations
\begin{equation*}
\gamma=\frac{1+\hat{c}_v}{\hat{c}_v} = \frac{2+D}{D}.
\end{equation*} 
 The equilibrium subsystem \eqref{eq:subgeneral} of the system of  ET$_{14}$ \eqref{eq:ET14} is the system of Euler equations.
The relationship between the unperturbed and perturbed states  is given by  the RH conditions \eqref{RHEq}  for the system of the Euler equations. 
Let $\mathbf{U}_0 = (\rho_0, v_0, T_0, 0, 0, 0)^T$ be the unperturbed state and the unperturbed Mach number $M_0$ is defined as 
\begin{equation*}
M_0 = \frac{s - v_0}{c_0}, 
\end{equation*}
where $c_0$ is the sound velocity in the unperturbed state. 
%%%%%%%%%%%%%%%
As is well known, except for contact shocks,  the solution of the RH equations for  Euler fluids is:
\begin{align}\label{6-43}
\begin{split}
&	V_1=V_{0} -\frac{2}{\gamma+1}\, V_{0}\,
	\dfrac{M_{0}^{2}-1}{M_{0}^{2}}, \qquad V=\frac{1}{\rho}, \\ 
& {v_1}={v}_{0}+ \frac{2c_{0}}{\gamma+1}\, \frac{M_{0}^{2}-1}{M_{0}}, \\
& T_1= T_{0}+2T_{0}\frac{(M_{0}^{2}-1)(\gamma
		M_{0}^{2}+1)(\gamma-1)}{M_{0}^{2}(1+\gamma)^{2}}. 
\end{split}
\end{align}%
It is also well known that we should take $M_0>1$ for obtaining the solution of a stable shock wave. 

Let us consider, without loss of generality, $v_0=0$ due to the Galilean invariance and let us define the dimensionless characteristic velocities as $\hat{\lambda} \equiv \lambda/c_0$. 
By considering only the two waves propagating in the positive $x$ directions and  taking \eqref{lame}, \eqref{soundv} and \eqref{6-43} into account, we obtain the dimensionless characteristic velocities in the unperturbed constant state $\mathbf{U}_0$ and in the perturbed constant state $\mathbf{U}_1$:
 \begin{align*}
 \begin{split}
&\hat{ \lambda}^{(1)}_0 = \Delta^{(1)}  , \qquad \hat{ \lambda}^{(2)}_0 = \Delta^{(2)}  \\
\\
&\hat{ \lambda}^{(1)}_1 = \frac{v_1}{c_0} + \frac{c_1}{c_0}\Delta^{(1)}  , \quad
 \hat{ \lambda}^{(2)}_1 = \frac{v_1}{c_0} + \frac{c_1}{c_0}\Delta^{(2)} .
 \end{split}
  \end{align*}
The former two are constant, while the latter two depend on $M_0$.
In the present case, the necessary condition for existence of sub-shock \eqref{necessaria}  expressed by the dimensionless variables reads:
\begin{equation}\label{neccond}
 \hat{ \lambda}_0 < M_0 < \hat{ \lambda}_1(M_0) < \hat{ \lambda}^{\max}_0.
\end{equation} 
Figure \ref{fig:ET14_diagram_D3} shows the dependence of the dimensionless characteristic velocities in the perturbed state $\hat{ \lambda}^{(1)}_1$ and  $\hat{\lambda}^{(2)}_1$ on the Mach number $M_0$ in the cases of $D=3$ (monatomic gas) and of $D=7$. 
It was proven that, in the limit of $D \rightarrow 3$, the solutions for rarefied polyatomic gases converge to the ones for rarefied monatomic gases  when we impose an appropriate initial condition, which is compatible with monatomic gases~\cite{singularlimit1,singularlimit2}. 

For $D=7$, we have $\hat{\lambda}^{(1)}_0 \simeq 0.773809$ and $\lambda^{(2)}_0 = \hat{\lambda}^{\max}_0 \simeq 1.74093$.
In contrast to the case $D=3$, for $D=7$, as we increase the Mach number from unity, the first characteristic velocity $\hat{ \lambda}^{(1)}_1$ evaluated  in the perturbed state $\mathbf{U}_1$   meets the shock velocity at $M_0 \simeq 1.31579$ before the fastest characteristic velocity in the unperturbed state.  
Therefore \eqref{neccond} is satisfied for $ 1.31579 < M_0 < 1.74093$ and, in principle, the sub-shock formation with smaller shock velocity than the maximum characteristic velocity may exist in this range.
However, as we will see in the next section, $M_0 = 1.31579$ is  a regular singular point and no sub-shock  arises until we reach $M_0 > 1.74093$, i.,e, until the shock velocity becomes larger than the maximum characteristic velocity evaluated in equilibrium state in front of the shock!
\subsection{Numerical analysis}
\label{sec:conjecture}
The shock structure was studied in \cite{ET14shock} for a non-polytropic rarefied gas by solving the ODE system \eqref{struttura} numerically for Mach numbers less than $1.47$ and the agreement between theoretical predictions and the experimental results is excellent with respect to previous theories.

In order to obtain the shock-structure solution also for large Mach number, in the present analysis, instead of solving the ODE system \eqref{struttura}, we use a different procedure solving ad hoc Riemann problem for the PDE  system \eqref{eq:ET14} 
according with the conjecture about the large-time behavior of the Riemann problem and the Riemann problem with structure \cite{Liu_struct} for a system of balance laws proposed by Ruggeri and coworkers \cite{Brini_Osaka,BriniRuggeri,MentrelliRuggeri} -- following an idea of  Liu \cite{Liu_conjecture}.
According to this conjecture, the solutions of both Riemann problems with and without structure, for large time, instead to converge  to the corresponding  Riemann problem of the equilibrium sub-system  (i.e combination of shock and rarefaction waves), converge   to  solutions that represent a  combination of shock structures (with and without sub-shocks) of the full system and rarefactions waves of the equilibrium subsystem. 

In particular, if the Riemann initial data correspond to a  shock family $\mathcal{S}$ of the equilibrium sub-system, for large time, the solution of the Riemann problem of the full system converges to the corresponding shock structure.
This means that, for the numerical study of the shock structure, instead of using a solver of ODE, which is not useful when a discontinuity (sub-shock) appears,  Riemann solvers (e.g. \cite{toro}) can be used and if we wait enough time after the initial time, we obtain the shock-structure profile with or without sub-shocks. 
This strategy was adopted in several shock phenomena  of ET \cite{RuggeriSugiyama}. 
In particular the conjecture was tested numerically for a Grad 13-moment system and a mixture of fluids \cite{Brini_Osaka, Brini_Wascom} and was verified in a simple $ 2\times 2$ dissipative model considered by Mentrelli and Ruggeri \cite{MentrelliRuggeri} for which it is possible to calculate the shock structures of the full system and  the rarefactions of the equilibrium subsystem analytically.

%%%%%%%%%%%%

We perform numerical calculations on the shock structure obtained after long time for the Riemann problem consisted with two equilibrium states $\mathbf{U}_0 = (\rho_0, 0, T_0, 0, 0, 0)^T$ and $\mathbf{U}_1 = (\rho_1, v_1, T_1, 0, 0, 0)^T$ satisfying the RH conditions for the system of the Euler equations \eqref{6-43}.
For the numerical calculations on the shock structure, the BGK model for the production terms is adopted and therefore the relaxation times $\tau_{\Pi}$, $\tau_{S}$ and $\tau_{q}$ are  constant and have the same value $\tau_{\Pi} = \tau_{S} = \tau_{q}$.  
We developed and adopted the parallel numerical code written in C language on the basis of the Uniformly accurate Central Scheme of order 2 (UCS2) proposed by Liotta, Romano and Russo~\cite{UCS2} for analyzing the hyperbolic balance laws with production term. 

Figure \ref{fig:rho_ET14} shows  typical examples of the mass density profile with $D=7$. 
The Mach numbers are $M_0=1.5$ and $M_0=2$. 
The profile for $M_0=1.5$ is continuous and no sub-shock arises. 
In the profile for $M_0=2$, only one sub-shock, which corresponds to the fastest mode, appears.  
The present situation is similar to the ones obtained in the case of a rarefied monatomic gas~\cite{Weiss,MullerRuggeri}. 
The singular point in the shock-structure solution becomes regular except for the maximum characteristic velocity. 
This result implies that the system of ET for rarefied polyatomic gas has the same property on the sub-shock formation and this property seems common for the systems satisfying the requirements of the ET theory.

\section{2 $\times$ 2 hyperbolic dissipative system}

\subsection{General form of 2 $\times$ 2 hyperbolic dissipative system}
Let us consider the following  $2\times 2$ dissipative  hyperbolic system of balance laws   proposed by Mentrelli and Ruggeri~\cite{MentrelliRuggeri}:
\begin{equation}
\begin{split}
&\frac{\partial u}{\partial t} + \frac{\partial}{\partial x}\left( \frac{\partial K}{\partial u} \right) = - \frac{1}{\tau}\left( u - v \right), \\
&\frac{\partial v}{\partial t} + \frac{\partial}{\partial x}\left( \frac{\partial K}{\partial v} \right) = - \frac{1}{\tau}\left( v - u \right), 
\label{eq:field_toy01}
\end{split}
\end{equation}
\begin{figure}[]
	\begin{center}
		\includegraphics[width=0.9\hsize] {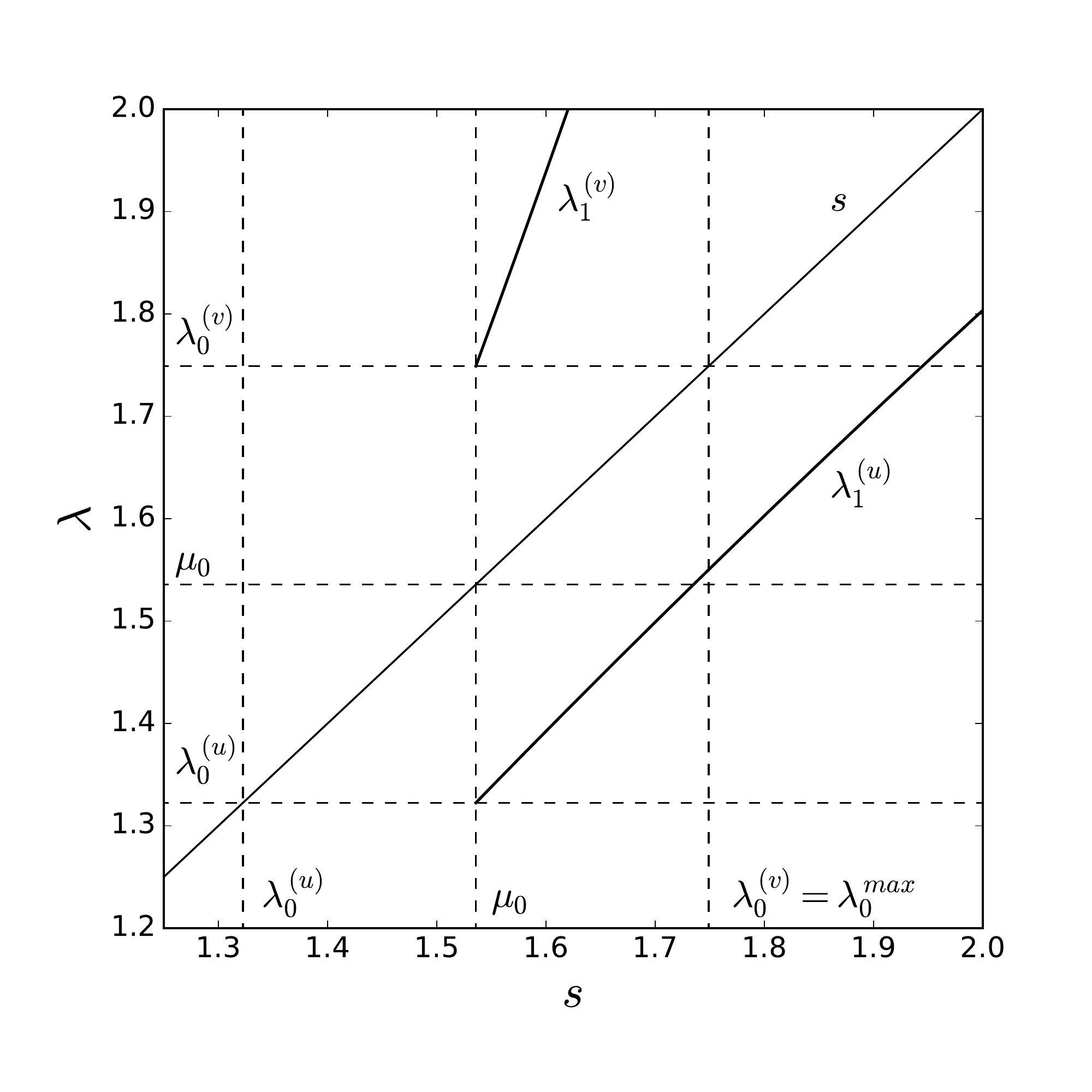}\\
		\caption{
			(Case A) Dependence of the characteristic velocities in the perturbed state $\lambda_1$ on the shock speed $s$ for $u_0=1.15$.
		}
		\label{diagram_u0-115}
	\end{center}
\end{figure}
or, alternatively, 
\begin{equation}
\begin{split}
&\frac{\partial }{\partial t}\left( u + v \right) + \frac{\partial }{\partial x}\left( \frac{\partial K}{\partial u} + \frac{\partial K}{\partial v} \right) = 0, \\
&\frac{\partial u}{\partial t} + \frac{\partial}{\partial x}\left( \frac{\partial K}{\partial u} \right) = - \frac{1}{\tau}\left( u - v \right)
\label{eq:field_toy02}
\end{split}
\end{equation}
for the unknown field $\mathbf{U}=(u,v)^T$, which is a function of space $x$ and time $t$.
Here $K \equiv K(u,v)$ is an arbitrary   smooth function in terms of the variables $u$ and $v$ and $\tau>0$ represents a  constant relaxation time.  
The equilibrium state is achieved when $u=v$. 

The system \eqref{eq:field_toy01} (or, \eqref{eq:field_toy02}) was proposed because this satisfies all the requirement of rational extended thermodynamics. 
In fact, the solution of the balance equations \eqref{eq:field_toy01} (or, \eqref{eq:field_toy02}) satisfies the following entropy inequality \cite{sign_entropy}:
\begin{equation}
\frac{\partial h}{\partial t} + \frac{\partial h^1}{\partial x} = \Sigma > 0, 
\label{eq:entropy_inequality}
\end{equation}
where $h$, $h^1$ and $\Sigma$ are, respectively,  the entropy density, the entropy flux and the entropy production density given by
\begin{equation}\label{entropie}
\begin{split}
&h = -\frac{1}{2}\left( u^2 + v^2 \right), \\
&h^1 = -u \frac{\partial K}{\partial u} - v \frac{\partial K}{\partial v} + K, \\
&\Sigma = \frac{1}{\tau} (u-v)^2. 
\end{split}
\end{equation}
Moreover, the concavity of the entropy density $h$ with respect to the field $(u,v)^T$ is automatically satisfied (see \eqref{entropie}$_1$). 

It is well known that, by introducing the formal substitution %chain rule
\begin{equation*}
    \partial_t \rightarrow -\lambda \delta \ \ \ \ \ \ \ \ \ \
    \partial_x \rightarrow \delta
\end{equation*}
and by putting zero for the production terms in \eqref{eq:field_toy01} (or \eqref{eq:field_toy02}), we obtain  a linear system of two equations where $\lambda$ represents the
characteristic velocity and $\left(\delta u,\delta v\right)^T$ is proportional to the characteristic eigenvector of the system 
associated with $\lambda$:
%that $\lambda$ are the characteristic velocity and $\uu$ are
%proportional to the characteristic eigenvectors. Therefore from
%\eqref{eq.toy.K1} or \eqref{eq.toy.K2}, we have:
%
\begin{equation} \label{eq.toy.lambda}
\begin{split}
    &\left(-\lambda + \frac{\partial^2 K}{\partial u^2}\right) \delta u + \frac{\partial^2 K}{\partial u \partial v}\delta v = 0, \\
    & \frac{\partial^2 K}{\partial u \partial v} \delta u + \left(-\lambda + \frac{\partial^2 K}{\partial u^2}\right) \delta v = 0.
\end{split}
\end{equation}
Therefore the characteristic velocities $\lambda^{(1)}$ and $\lambda^{(2)}$ are obtained as the solutions of the characteristic polynomial $P(\lambda) = 0$, where
\begin{equation*}
%\begin{split}
P(\lambda)  = \lambda^2 - \left\{\frac{\partial^2 K}{\partial u^2}  +  \frac{\partial^2 K}{\partial v^2}\right\} \lambda  + \frac{\partial^2 K}{\partial u^2}  \frac{\partial^2 K}{\partial v^2} -\left(\frac{\partial^2 K}{\partial u \partial v} \right)^2. 
%\end{split}
\end{equation*}
In particular, the equilibrium characteristic velocities $\lambda^{(1)}_E$ and $ \lambda^{(2)}_E$ are the roots of  $P_E(\lambda_E) = 0$, where
\begin{equation}\label{PElambda}
\begin{split}
P_E(\lambda_E) &= \lambda_E^2 -\left. \left\{\frac{\partial^2 K}{\partial u^2}  +  \frac{\partial^2 K}{\partial v^2}\right\}\right|_E \lambda_E  +\\
&\left.\left\{ \frac{\partial^2 K}{\partial u^2}  \frac{\partial^2 K}{\partial v^2} -\left(\frac{\partial^2 K}{\partial u \partial v} \right)^2\right\}\right|_E. 
\end{split}
\end{equation}
Here the quantities with subscript $E$ represent the quantities evaluated in the equilibrium state in which $v=u$.

According with the definition   given by Boillat and Ruggeri \cite{SubSystem}, in the present case, the equilibrium subsystem associated with the system \eqref{eq:field_toy02} is obtained, by putting $v=u$ into the equation \eqref{eq:field_toy02}$_1$:
\begin{equation}\label{equilibriumss}
\frac{\partial u}{\partial t} + \frac{1}{2}\frac{\partial}{\partial x}\left(\frac{d \bar{K}}{d u}\right) = 0, 
\end{equation}
where $\bar{K} = \bar{K}(u)$ is defined by $\bar{K}(u) = K (u,u)$. 
The characteristic velocity $\mu$  of the equilibrium subsystem \eqref{equilibriumss} is given by 
\begin{equation}\label{mumu}
\mu = \frac{1}{2}\frac{d^2 \bar{K}}{d u^2}. 
\end{equation}
\begin{figure}[]
	\begin{center}
		\includegraphics[width=0.9\hsize] {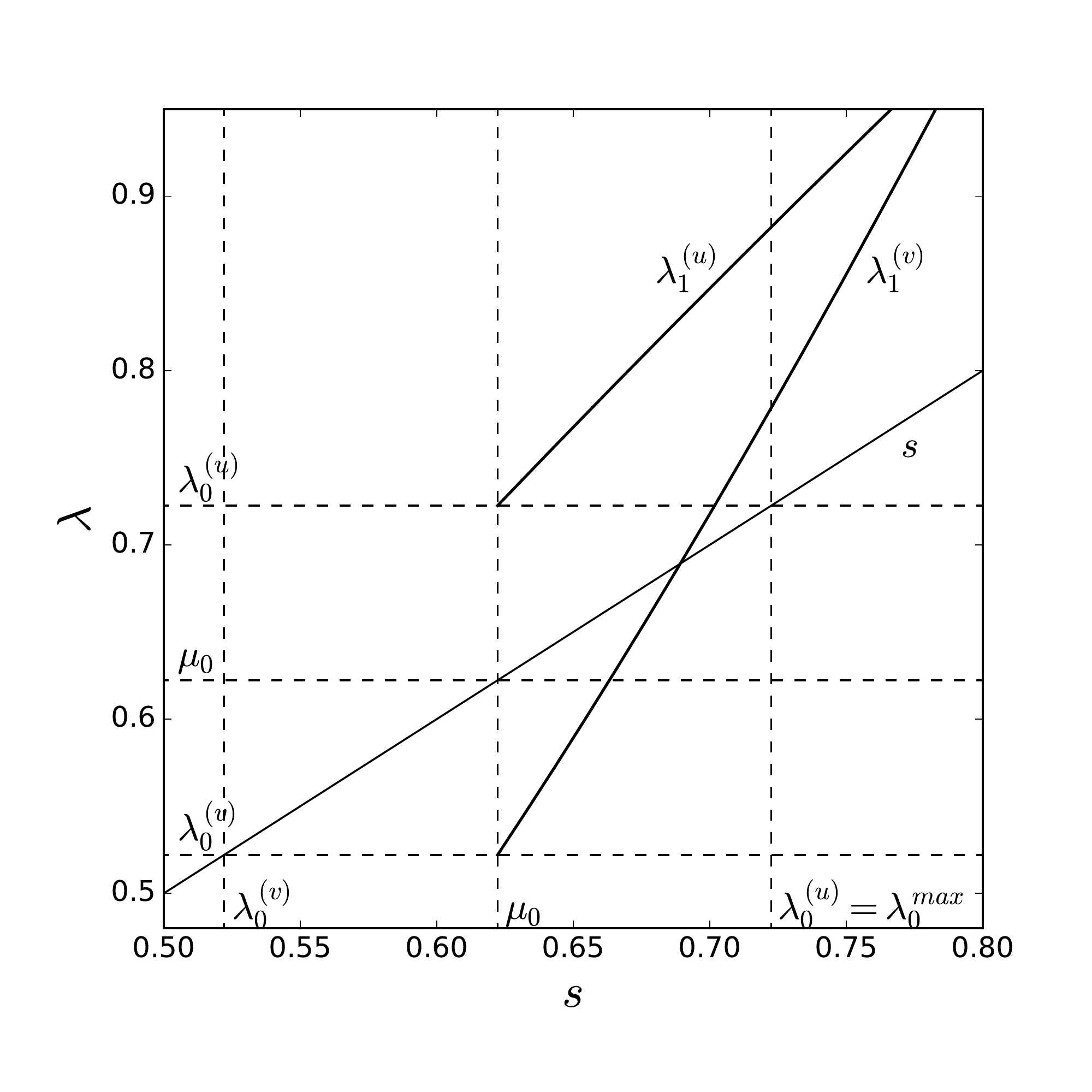}
		\caption{
			(Case B) Dependence of the characteristic velocities in the perturbed state $\lambda_1$ on the shock speed $s$ for $u_0=0.85$.
		}
		\label{fig:toy_diagram_u0-085}
	\end{center}
\end{figure}
Taking into account the following identities:
\begin{align*}
& \frac{d \bar{K}}{du} = \left.\left( \frac{\partial K}{\partial u} + \frac{\partial K}{\partial v} \right)\right|_E, \\
& \frac{d^2 \bar{K}}{du^2} = \left.\left( \frac{\partial^2 K}{\partial u^2} + 2\frac{\partial^2 K}{\partial u \partial v}+ \frac{\partial^2 K}{\partial v^2} \right)\right|_E, \\
\end{align*}
we have 
\begin{equation}\label{12b}
P_E(\mu) = - \frac{1}{4}\left\{\left.\left(  \frac{\partial^2 K}{\partial u^2}  -  \frac{\partial^2 K}{\partial v^2} \right)\right|_E\right\}^2  \leq 0.  
\end{equation}
Therefore,  we have the sub-characteristic conditions~\cite{SubSystem}:
\begin{equation} \label{13b}
\lambda^{(1)}_E \leq \mu \leq \lambda^{(2)}_E.
\end{equation}
The system \eqref{eq:field_toy02} also belongs to the general hyperbolic system of balance laws in one-space dimension \eqref{sistemagenerale} with
\begin{align}\label{sistg}
\begin{split}
&\mathbf{U} \equiv (u+v,u)^T, \quad \mathbf{F} \equiv \left(\left( \frac{\partial K}{\partial u} + \frac{\partial K}{\partial v} \right),\frac{\partial K}{\partial u}\right)^T, \\
& \mathbf{f} \equiv \left(0, -\frac{1}{\tau}(u-v)\right)^T.
\end{split}
\end{align}
This kind of dissipative hyperbolic systems have recently been studied with particular attention to the existence of global smooth solutions. In fact, under the Shizuta-Kawashima coupling condition (K-condition) \cite{Kaw-1985,Kaw-1987}
\begin{equation}\label{Kc}
    \nabla \mathbf{f} \cdot \mathbf{r}^{(i)}\Big|_E \neq 0 \ \ \ \ \ \
    \forall \ i=1,\ldots,N, 
\end{equation}
($\mathbf{r}^{(i)}$ represents the $i^{th}$ characteristic eigenvector of the hyperbolic system \eqref{sistemagenerale}), it was proven that, for small initial data, smooth solutions exist for all times and
constant states are stable \cite{Nat-2003,Yong,RugSerre,Nat-2005}.
The K-condition \eqref{Kc} is equivalent to \cite{TR.kaw}:
\begin{equation*} \label{eq.Kcondition}
    \delta \mathbf{f} \Big|_E \neq 0.
\end{equation*} 
In the present case, from \eqref{sistg}$_3$, we have
\begin{equation}\label{dudv}
 \delta \mathbf{f}|_E \neq 0 \quad \Longleftrightarrow \quad  (\delta u -\delta v)|_E \neq 0.
\end{equation}
We need to consider the two possible cases separately:
\begin{itemize}
\item If
\begin{equation}\label{caso1}
\left. \frac{\partial^2 K}{\partial u \partial v}  \right|_E =0, 
\end{equation}
from \eqref{PElambda} and \eqref{eq.toy.lambda}, we have
 \begin{align*}
& \lambda^{(1)}_E= \left.\frac{\partial^2 K}{\partial u^2}\right|_E , \quad (\delta u)|_E= 1, \,\, ( \delta v)|_E= 0, \\
& \lambda^{(2)}_E= \left. \frac{\partial^2 K}{\partial v^2}\right|_E  , \quad (\delta u)|_E= 0, \,\, ( \delta v)|_E= 1
 \end{align*}
and  \eqref{dudv} is automatically satisfied for both eigenvectors. 
\item If
\begin{equation*}
\left. \frac{\partial^2 K}{\partial u \partial v}  \right|_E \neq 0,
\end{equation*}
 from \eqref{eq.toy.lambda}, we have
\begin{equation*}
(\delta u)|_E= - \left.\frac{\partial^2 K}{\partial u \partial v} \right|_E, \qquad 
(\delta v)|_E= -\lambda_E +\left.\frac{\partial^2 K}{\partial u^2} \right|_E
\end{equation*}
and therefore the K-condition \eqref{dudv} is satisfied when 
\begin{equation}\label{omega}
\lambda_E \neq \omega, \quad \text{with} \quad \omega = \left.\frac{\partial^2 K}{\partial u^2} \right|_E + \left.\frac{\partial^2 K}{\partial u \partial v} \right|_E.
\end{equation}
From \eqref{PElambda}, we  have 
\begin{equation*}
P_E(\omega) = \left.\frac{\partial^2 K}{\partial u \partial v} \right|_E \left.\left(  \frac{\partial^2 K}{\partial u^2}  -  \frac{\partial^2 K}{\partial v^2} \right)\right|_E. 
\end{equation*}
The K-condition  \eqref{omega} implies $P_E(\omega) \neq 0$ and therefore
\begin{equation}\label{eq:SKC}
\left.\left( \frac{\partial^2 K}{\partial u^2}  -  \frac{\partial^2 K}{\partial v^2} \right)\right|_E \neq 0. 
\end{equation}
We notice that, if \eqref{eq:SKC} holds,   the equilibrium characteristic velocities for the full system have the different values from the one for the equilibrium subsystem  and the inequalities in \eqref{12b} and \eqref{13b} become strict.
\end{itemize}
\begin{figure}[]
	\begin{center}
		\includegraphics[width=0.9\hsize] {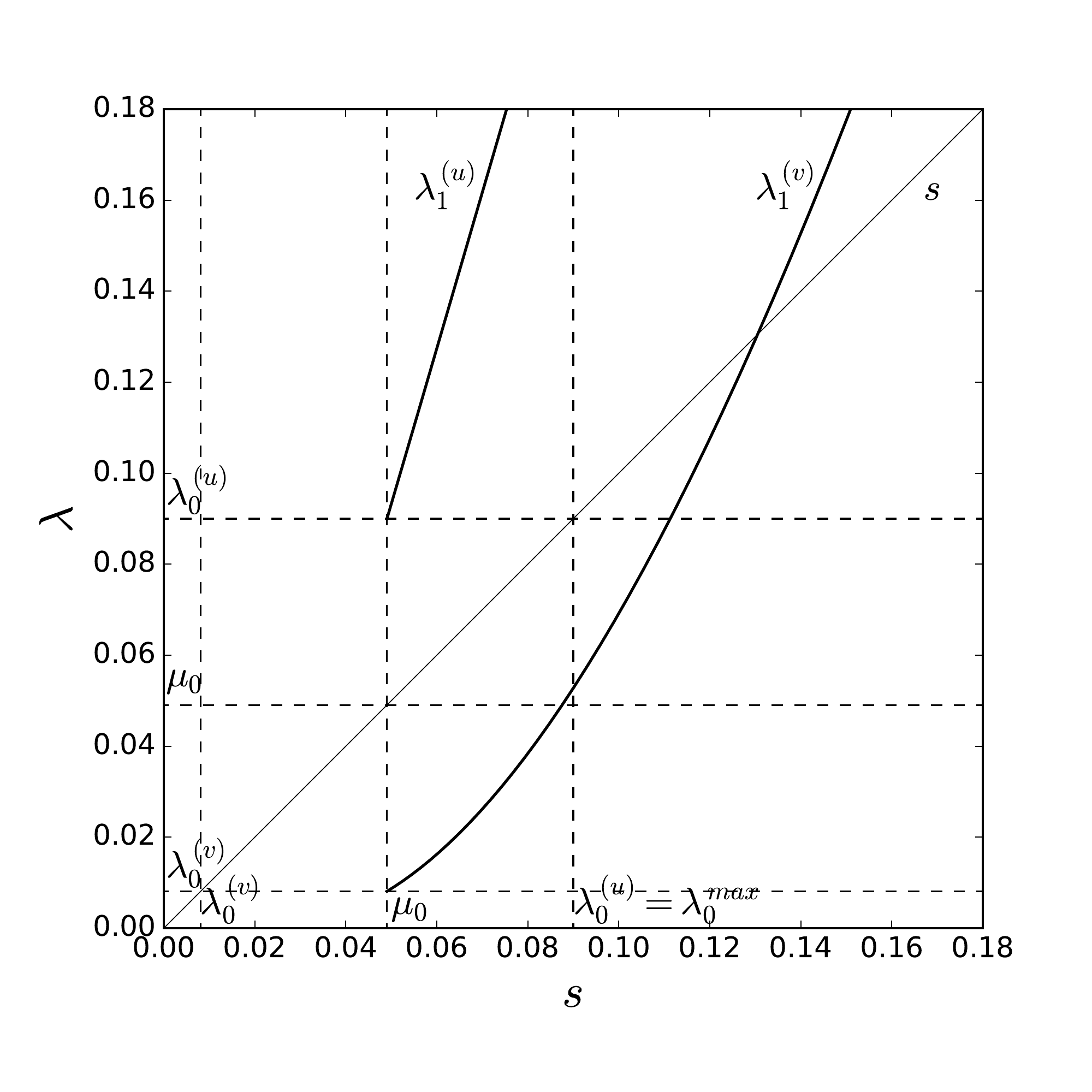}\\
		\caption{
			(Case C) Dependence of the characteristic velocities in the perturbed state $\lambda_1$ on the shock speed $s$ for $u_0=0.3$.
		}
		\label{fig:toy_diagram_u0-03}
	\end{center}
\end{figure}
Therefore we can summarize as follows: 
\begin{theorem}
For any smooth function $K(u,v)$ such that 
\begin{align*}
& \left. \frac{\partial^2 K}{\partial u \partial v}  \right|_{u=v} =0,
\\
& \text{or} \\
& \left.\left( \frac{\partial^2 K}{\partial u^2}  -  \frac{\partial^2 K}{\partial v^2} \right)\right|_{u=v} \neq 0,
\end{align*}	
and initial data sufficiently small, according with the theorems stated in{ \rm{\cite{Nat-2003,Yong,RugSerre,Nat-2005}}}, the system \eqref{eq:field_toy01} has global smooth solutions for all time.
\end{theorem}

\subsection{2 $\times$ 2 system with $K=u^4/12+v^6/30$}
In the paper   \cite{MentrelliRuggeri}, the system with $K = u v^2$ was studied. 
In this system, two characteristic velocities have the different sign; one is positive and another one is negative. 
In order to discuss the sub-shock formation with the slower shock velocity than the maximum characteristic velocity, we need to construct a new system in which both characteristic velocities are positive. 
In the present paper, we adopt
\begin{equation*}
K=u^4/12+v^6/30. 
\end{equation*}
In this case, we have the following balance equations: 
\begin{equation}
\begin{split}
&\frac{\partial u}{\partial t} + \frac{\partial}{\partial x}\left( \frac{u^3}{3} \right) = - \frac{u-v}{\tau}, \\
&\frac{\partial v}{\partial t} + \frac{\partial}{\partial x}\left( \frac{v^5}{5} \right) = - \frac{v-u}{\tau}, \label{eq:field_toy1}
\end{split}
\end{equation}
or, alternatively, 
\begin{equation}
\begin{split}
&\frac{\partial}{\partial t}\left( u + v \right) + \frac{\partial}{\partial x}\left( \frac{u^3}{3} + \frac{v^5}{5} \right) = 0, \\
&\frac{\partial u}{\partial t} + \frac{\partial}{\partial x}\left( \frac{u^3}{3} \right) = - \frac{u-v}{\tau}. 
\label{eq:field_toy2}
\end{split}
\end{equation}
The solution of the balance equations \eqref{eq:field_toy1} (or, \eqref{eq:field_toy2}) satisfies the entropy inequality \eqref{eq:entropy_inequality} where
the entropy density $h$, the entropy flux $h^1$ and the entropy production density $\Sigma$ are, in the present case, given by
\begin{equation*}
\begin{split}
&h = -\frac{1}{2}\left( u^2 + v^2 \right), \\
&h^1 = - \frac{u^4}{4} - \frac{v^6}{6}, \\
&\Sigma = \frac{1}{\tau} (u-v)^2. 
\end{split}
\end{equation*}
The characteristic velocities $\lambda$ are 
\begin{equation}
\lambda = u^2, v^4. 
\label{eq:toy_lambda}
\end{equation}
We adopt the following notation:
\begin{equation*}
\lambda^{(u)}=u^2 \quad \text{and} \quad \lambda^{(v)}=v^4. 
\end{equation*}
The equilibrium sub-system \eqref{equilibriumss} becomes 
\begin{equation}\label{eq:toy-eqsub}
\frac{\partial u}{\partial t} + \frac{\partial }{\partial x}  \left( \frac{u^3}{6} + \frac{u^5}{10} \right) = 0 
\end{equation}
and the characteristic velocity of the equilibrium sub-system $\mu$ \eqref{mumu} becomes 
\begin{equation}
\mu = \frac{u^2+u^4}{2}. 
\label{eq:toy_mu}
\end{equation}
From equations \eqref{eq:toy_lambda} and \eqref{eq:toy_mu}, it can be easily proven that the sub-characteristic condition \eqref{13b}  holds.
The Shizuta-Kawashima condition is always satisfied because the condition  \eqref{caso1} holds.

\section{Identification of possible sub-shocks}
Let us consider the shock-structure solution of  the system \eqref{eq:field_toy2}. 
In the present case, \eqref{eq:shockstructure_general} becomes:
\begin{align}\label{RHsub}
\begin{split}
& \frac{d}{dz}\left\{-s(u+v)+\left(\frac{u^3}{3} +\frac{v^5}{5}
\right)\right\} = 0, 
 \\
& \left(-s + u^2\right)\frac{d u}{dz}=\frac{v-u}{\tau},
\end{split}
\end{align}
with the following boundary conditions
\begin{equation}\label{contorno2}
\lim_{z\rightarrow  + \infty  } (u,v)=(u_0,u_0), \qquad \lim_{z \rightarrow  - \infty } (u,v)=(u_1,u_1). 
\end{equation}
From \eqref{RHsub}$_1$ we have :
\begin{equation*}
-s(u+v)+\left(\frac{u^3}{3} +\frac{v^5}{5}
\right) = \text{const.}
\end{equation*}
and by taking \eqref{contorno2} into account, we obtain  the Rankine-Hugoniot conditions for the equilibrium subsystem  \eqref{eq:toy-eqsub}:
\begin{equation}\label{RH2}
-2 s u_0+\left(\frac{u_0^3}{3} +\frac{u_0^5}{5}
\right) = -2 s u_1+\left(\frac{u_1^3}{3} +\frac{u_1^5}{5} \right) = \text{const.}
\end{equation}
\begin{figure*}[]
	\begin{center}
		\includegraphics[width=0.32\hsize] {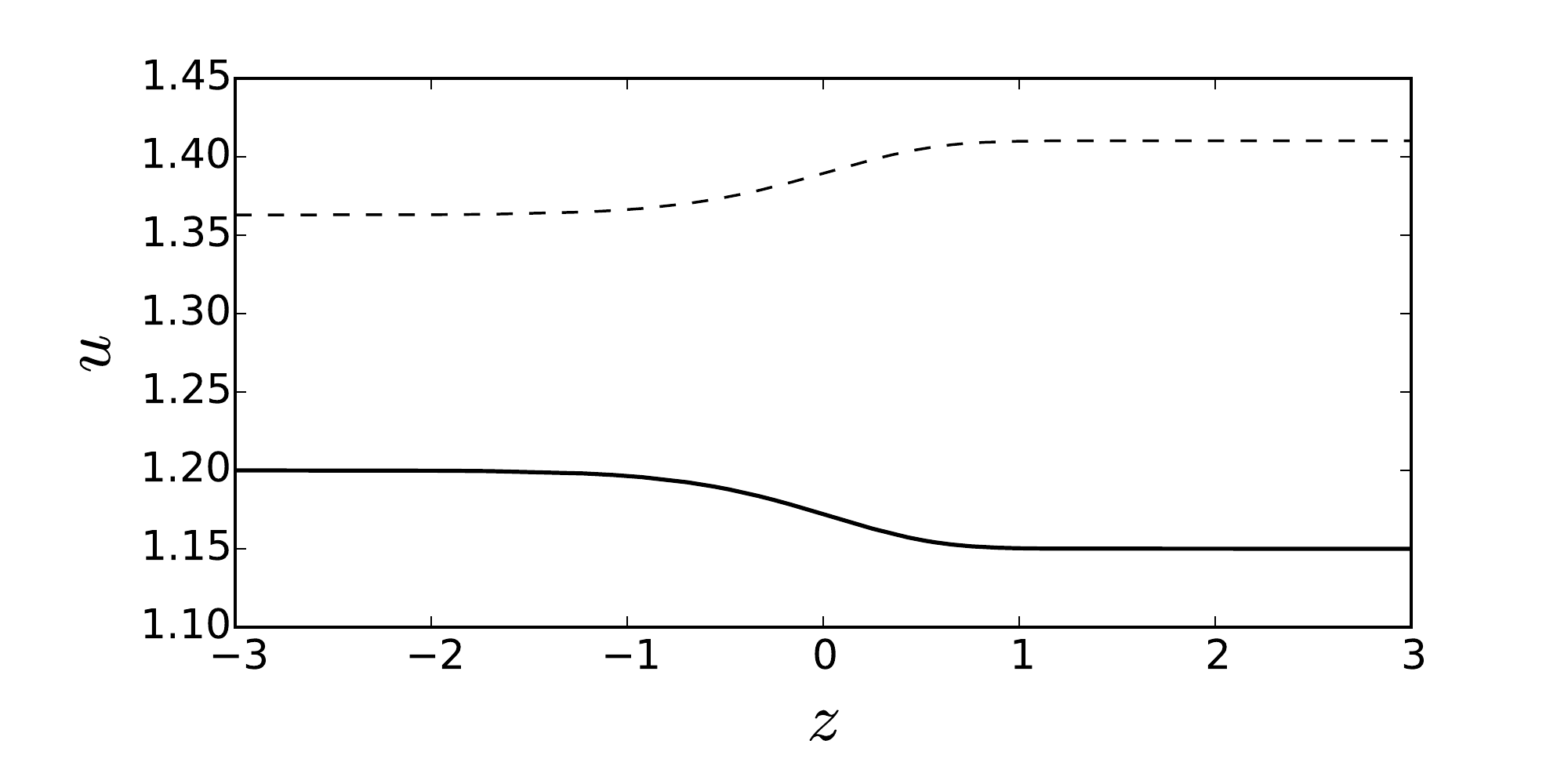}
		\includegraphics[width=0.32\hsize] {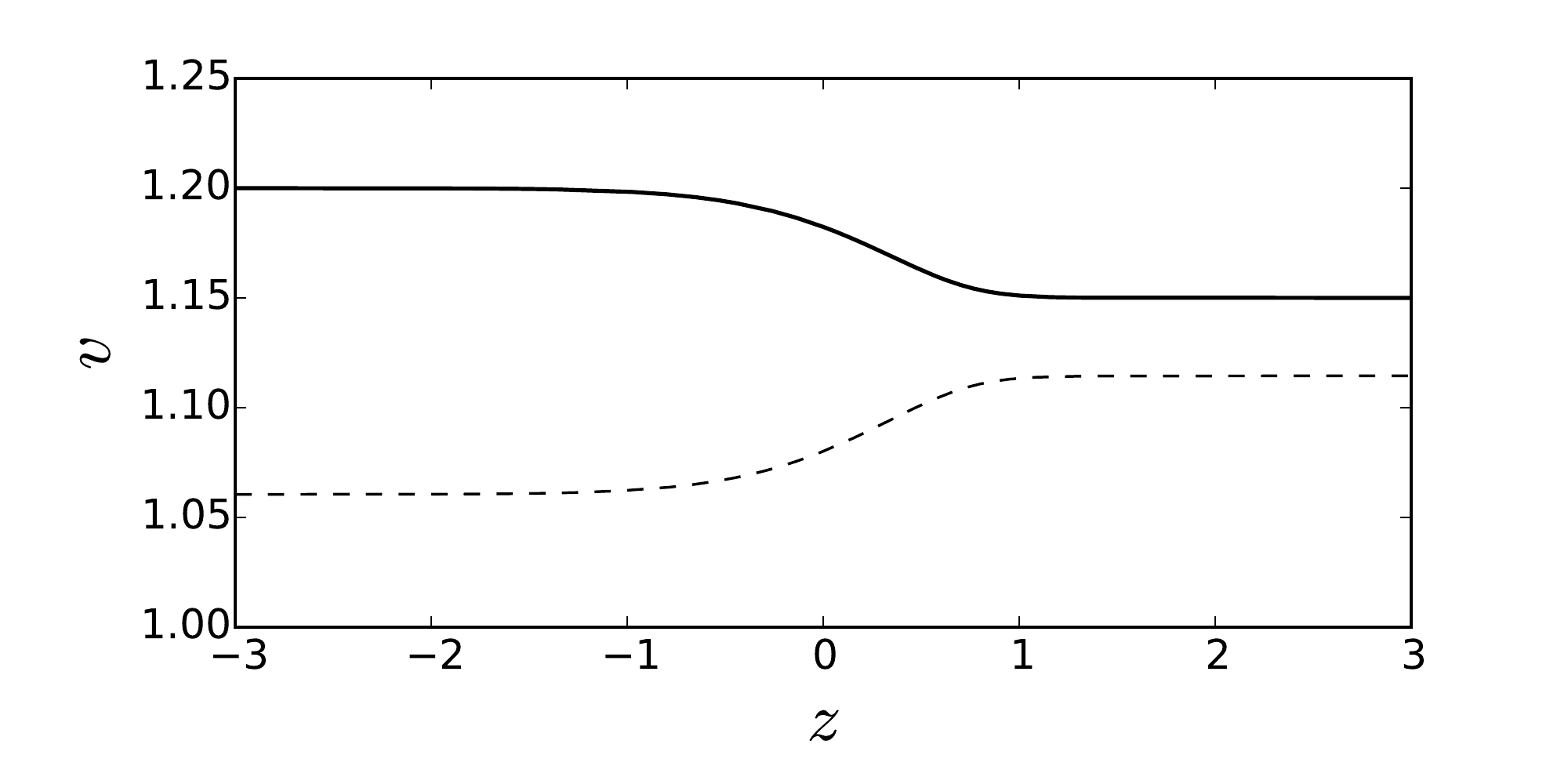}
		\includegraphics[width=0.32\hsize] {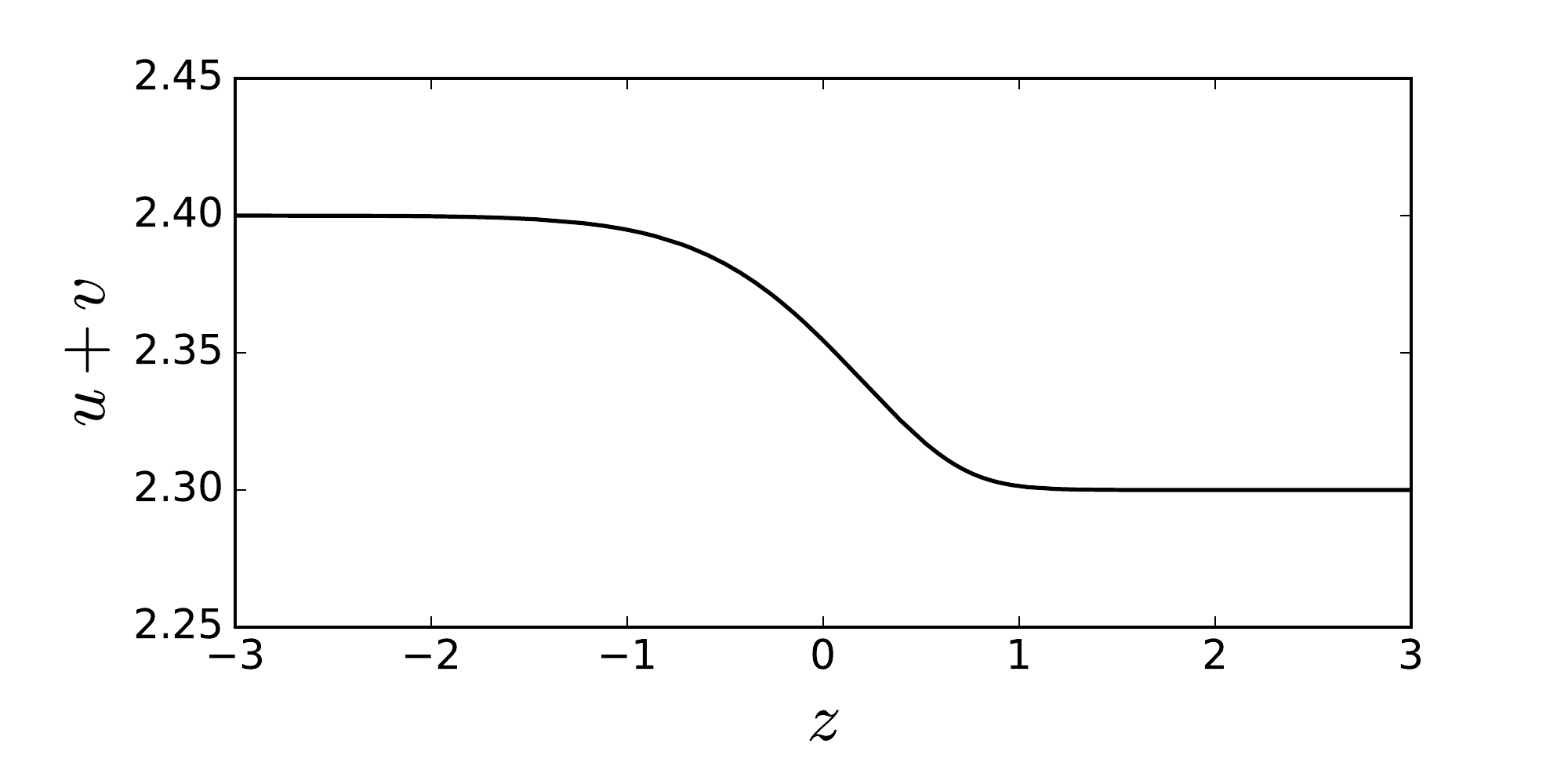}
		\includegraphics[width=0.32\hsize] {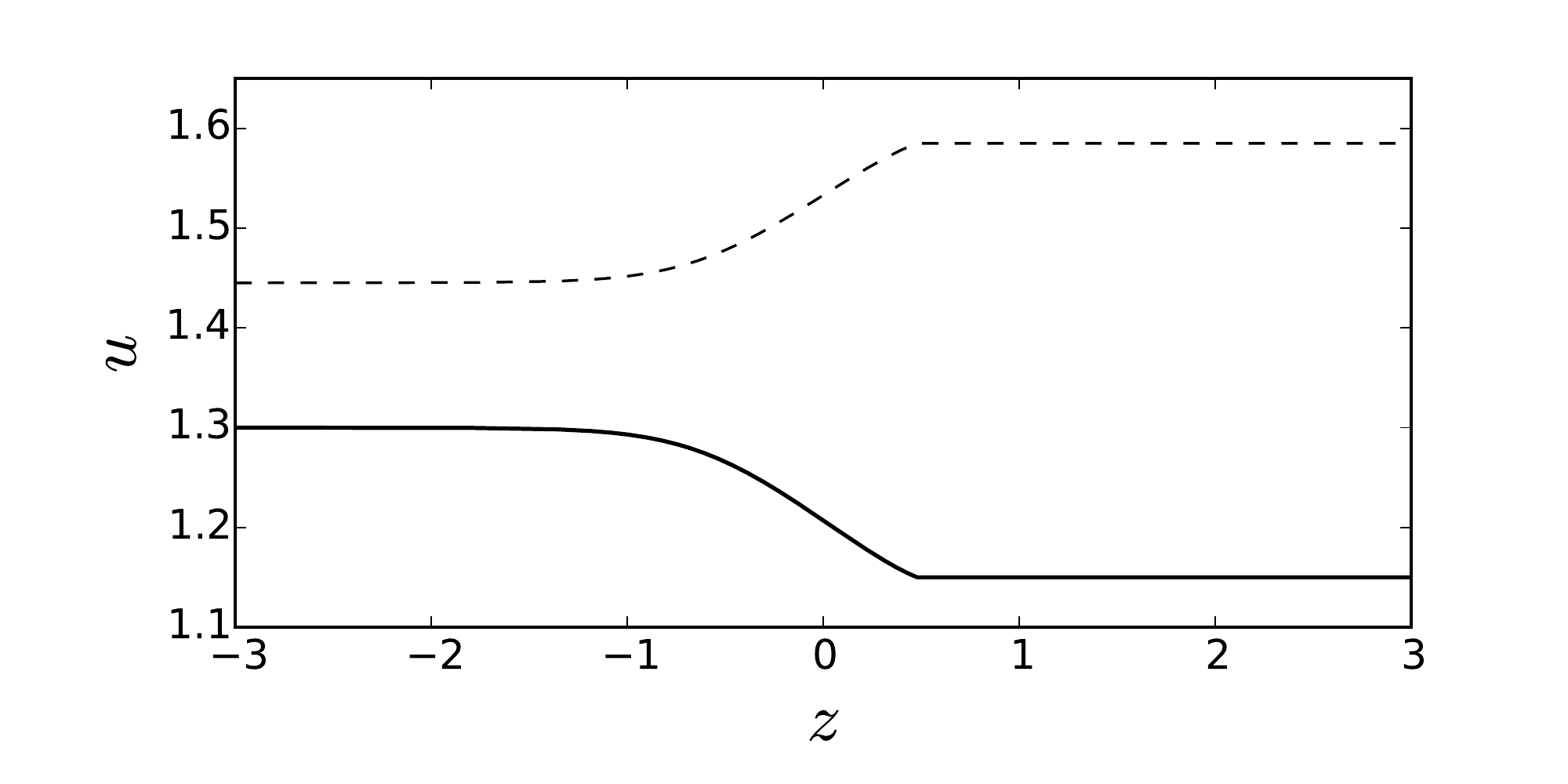}
		\includegraphics[width=0.32\hsize] {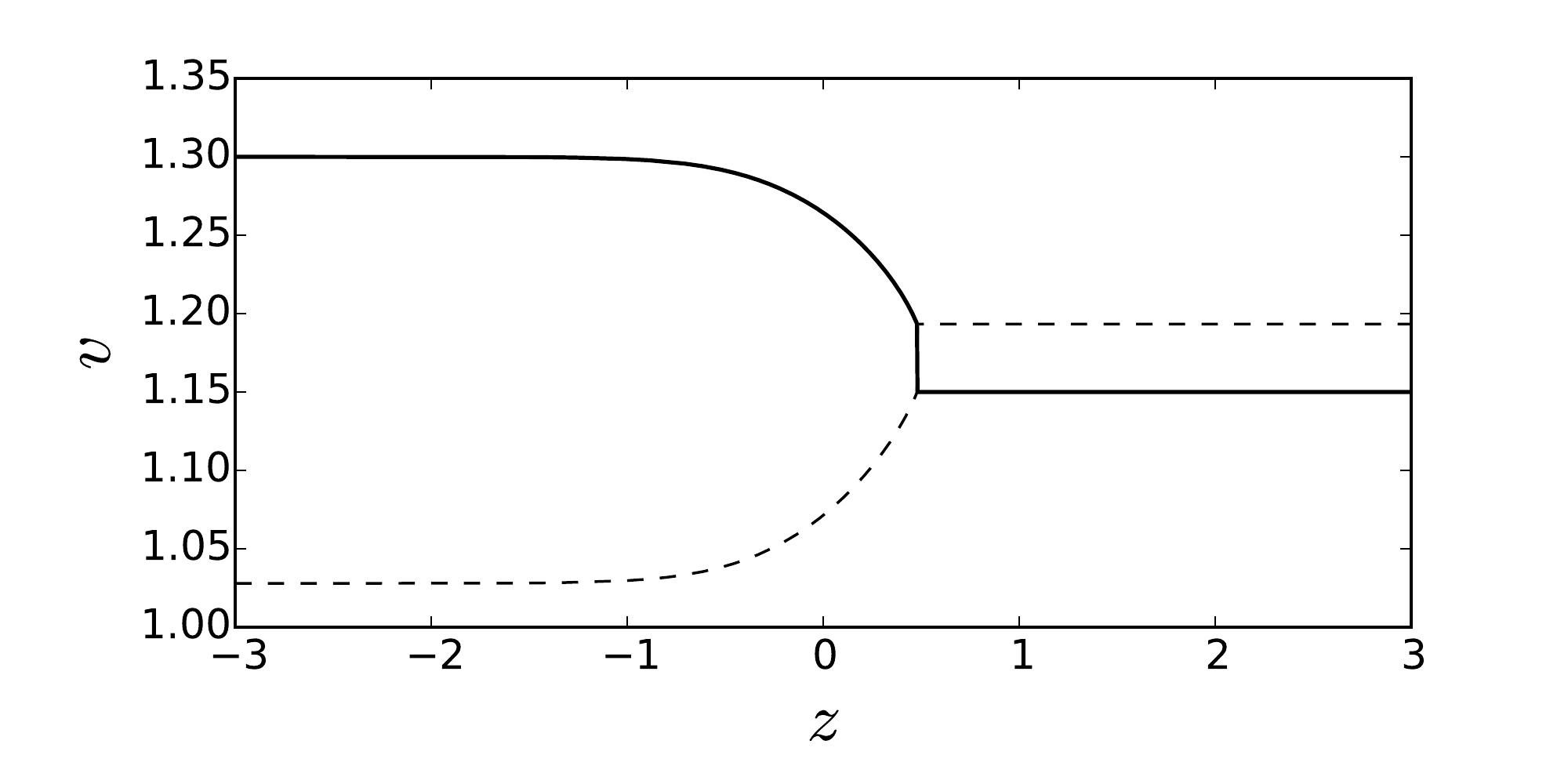}
		\includegraphics[width=0.32\hsize] {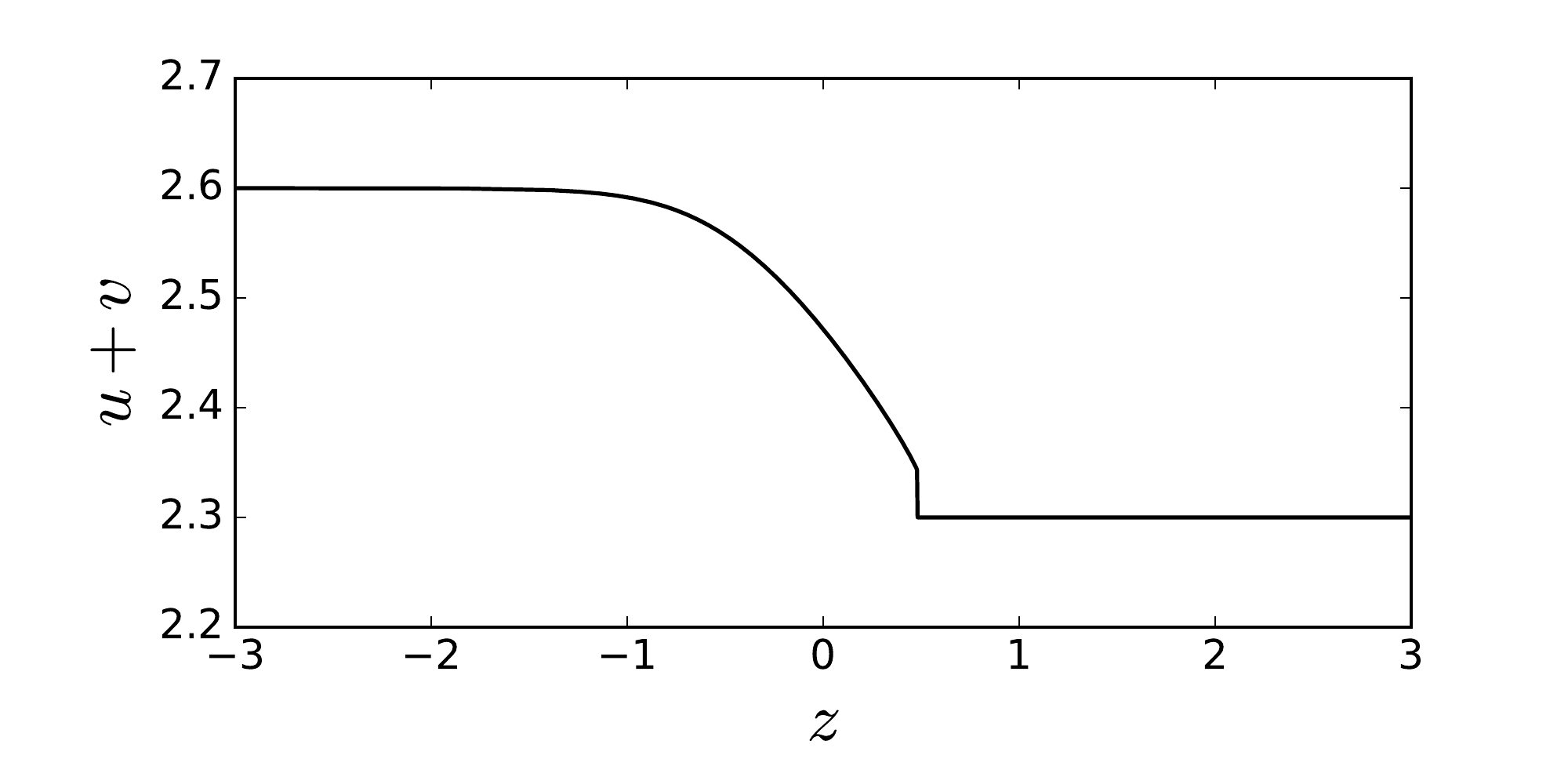}
		\caption{
			(Case A) Shock structure (solid curves) for $u_1 = 1.2$ (top) and $u_1 = 1.3$ (bottom). 
			The possible state just after the sub-shock (dotted curves) predicted by the RH conditions is also shown. 
			$u_0 = 1.15$ and $\tau = 1$. 
		}
		\label{fig:u_u0-115_u1-12}
	\end{center}
\end{figure*}
Therefore we have a relation between $s$, $u_0$ and $u_1$ given by \eqref{RH2}, and, by excluding the null shock $u_1=u_0$, the relation can be rewritten as:
\begin{equation}
s =\frac{u_1^2+u_0 u_1 + u_0^2}{6} + \frac{u_1^4+u_0 u_1^3+u_0^2u_1^2+u_0^3 u_1+u_0^4}{10}. 
\label{eq:RH_toy_eq} 
\end{equation}
From the RH conditions and the expression of the characteristic eigenvalue of the subsystem \eqref{eq:toy_mu}, we conclude that the Lax condition~\cite{Lax} for the equilibrium subsystem is satisfied when
\begin{equation*}
\mu_0 < s < \mu_1,  \quad \text{provided}  \quad u_1>u_0>0. 
\label{eq:lax_subsystem}
\end{equation*}
In the present case we have:
\begin{align*}
\begin{split}
& \lambda^{(u)}_0 = u_0^2, \qquad  \lambda^{(v)}_0 = u_0^4 \\
& \lambda^{(u)}_1 = u_1^2, \qquad  \lambda^{(v)}_0 = u_1^4.
\end{split}
\end{align*}
The first two are constants depending on $u_0$ and the other two are functions of $s$ through 
the relation \eqref{eq:RH_toy_eq} that gives $u_1$ as function of $s$ and $u_0$.

We classify the RH curves into the following three different cases:
Case. A: $u_0>1.06$; Case. B:   $0.536<u_0 <1$; Case. C: $0<u_0 <0.536$ or $1<u_0 <1.06$.

\subsection{Case A}

If we choose the unperturbed state $\mathbf{U}_0 = (u_0, u_0)^T$ with  $u_0>1.06$, the relationship $\lambda^{(u)}_0 < \lambda^{(v)}_0$ holds and both characteristic velocities in the perturbed state $\mathbf{U}_1 = (u_1, u_1)^T$,  $\lambda^{(u)}_1<s$ and $\lambda^{(v)}_1>s$  never meet the shock velocity $s$.
 Therefore the necessary conditions \eqref{necessaria} are  violated and there exists only one possibility of sub-shock formation when the shock velocity is larger than the maximum characteristic velocity: $s > \lambda^{(v)}_0$. 
As a typical example, we show the shock velocity dependence of the characteristic velocities in the perturbed state $\mathbf{U}_1 = (u_1, u_1)^T$ for $u_0=1.15$ in Figure \ref{diagram_u0-115}. 
In this case, $\mu_0 \simeq 1.54$, $\lambda^{(1)}_0 \simeq 1.32$ and $\lambda^{(2)}_0 \simeq 1.75$.

\subsection{Case B}
If we choose  $\mathbf{U}_0 = (u_0, u_0)^T$  with  $0.536<u_0<1$, the relationship $\lambda^{(u)}_0 > \lambda^{(v)}_0$ holds. 
The characteristic velocity   $\lambda^{(v)}_1$ in the perturbed state $\mathbf{U}_1 = (u_1, u_1)^T$ is equal to the shock velocity at the critical characteristic velocity $s_*$, which is smaller than the maximum characteristic velocity in the unperturbed state; $s_*<\lambda^{(u)}_0$. 
There are two possibilities of the sub-shock formation. 
The first possibility is the sub-shock when $s_*<s $. 
The second  is the sub-shock when $s > \lambda^{(u)}_0$. 
The necessary condition \eqref{necessaria} holds for $s_* < s < \lambda^{(u)}_0$. 
Therefore this case is a candidate of a counter example to have a sub-shock with the shock velocity smaller than the maximum characteristic velocity in the unperturbed state and also to have multiple sub-shock. 
As a typical example, we show the shock velocity dependence of the characteristic velocities in the perturbed state for $u_0=0.85$ in Figure \ref{fig:toy_diagram_u0-085}. 
In the present case, $\mu_0=0.622$, $\lambda^{(u)}_0=0.723$, $\lambda^{(v)}_0=0.522$ and $s_*=0.689$. 
\begin{figure*}[]
	\begin{center}
		\includegraphics[width=0.32\hsize] {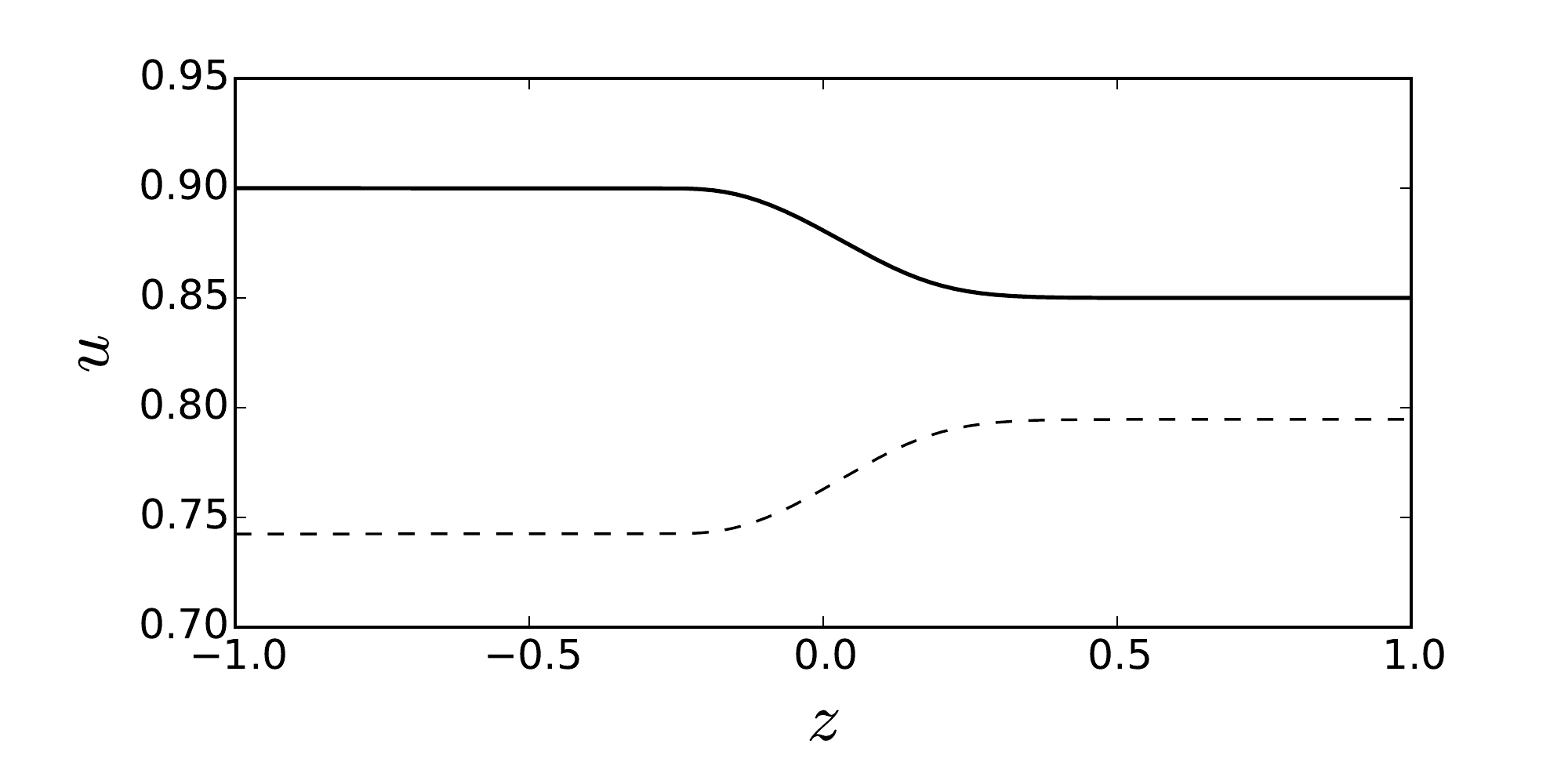}
		\includegraphics[width=0.32\hsize] {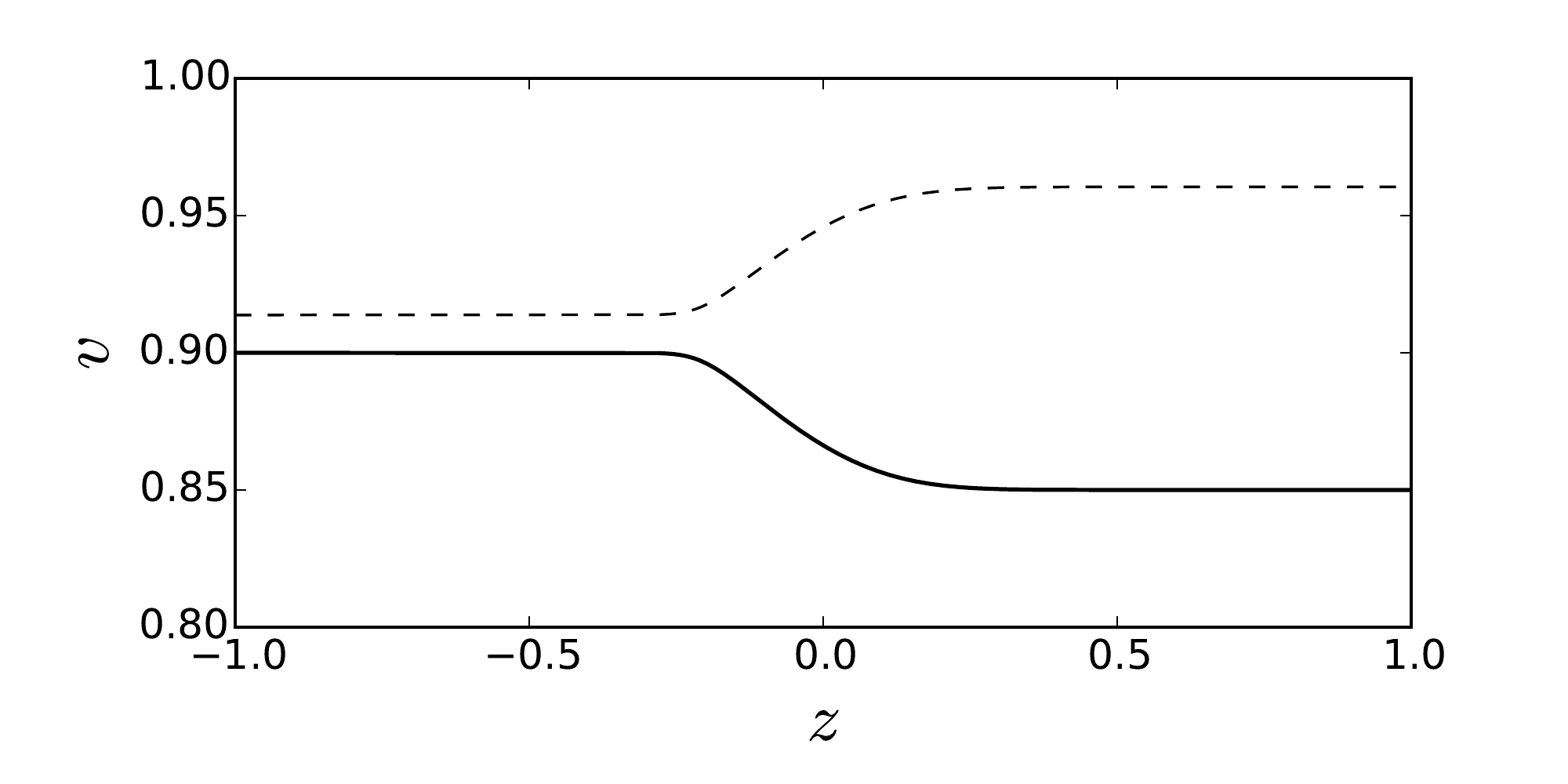}
		\includegraphics[width=0.32\hsize] {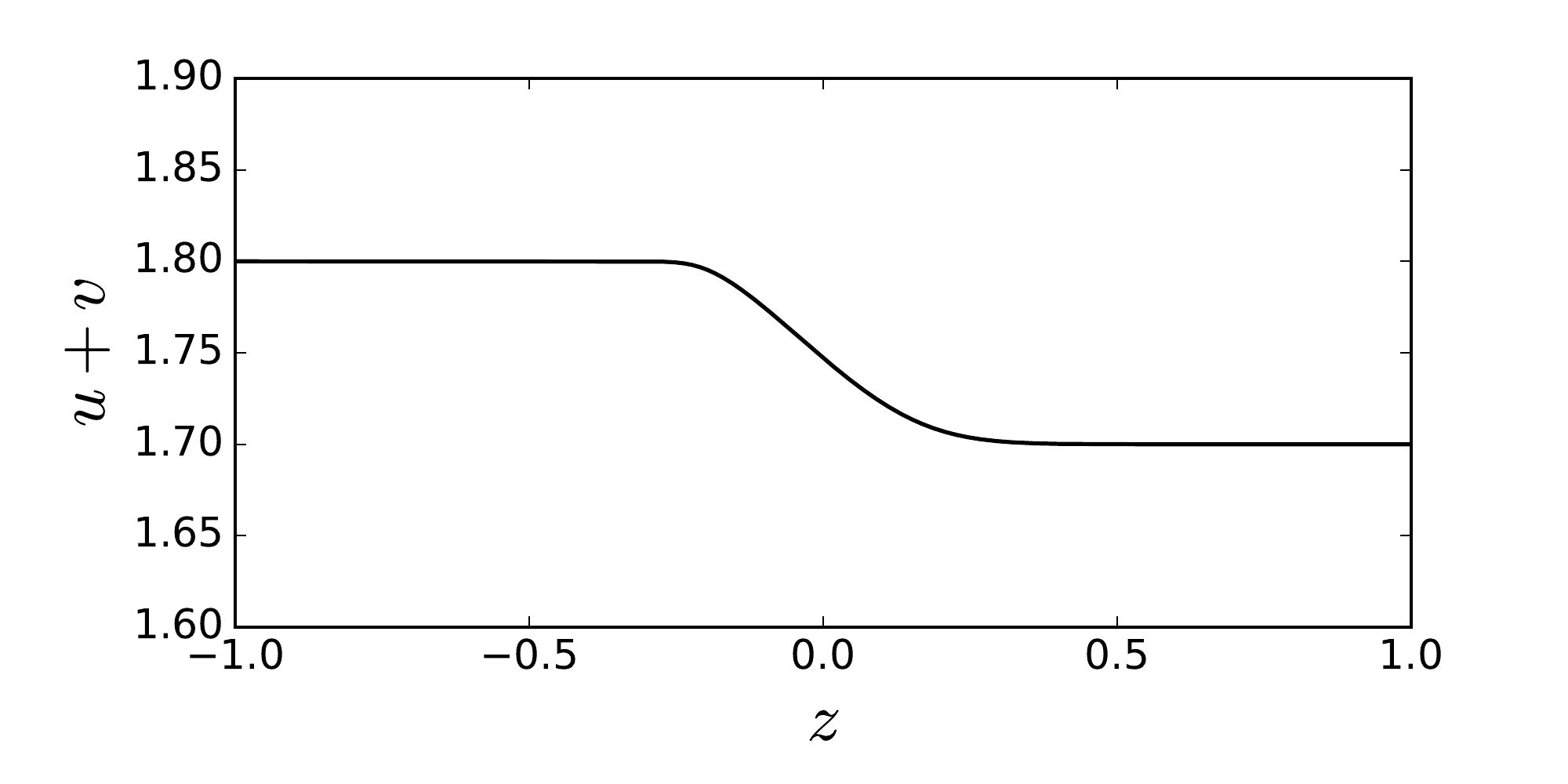}
		\includegraphics[width=0.32\hsize] {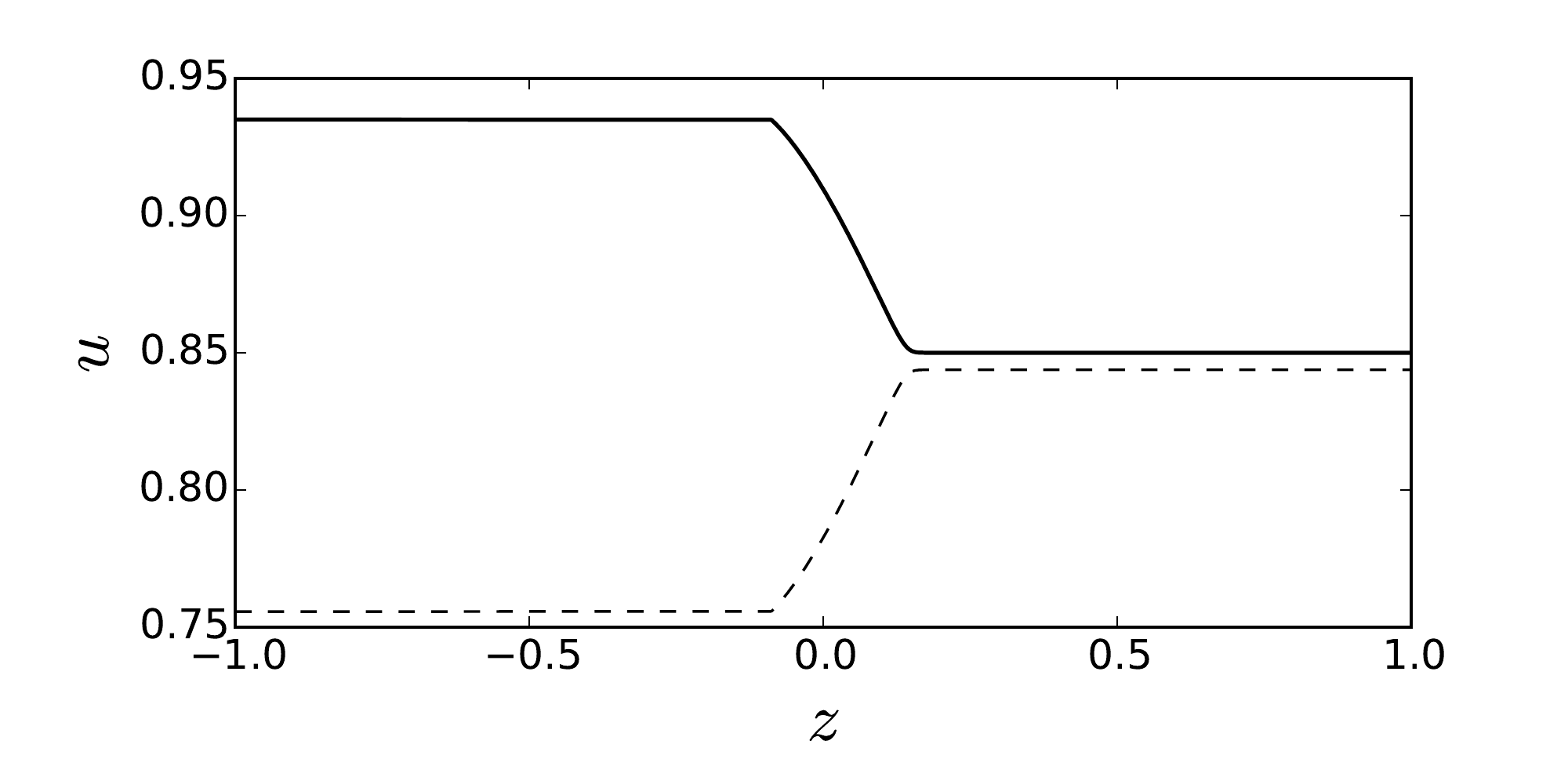}
		\includegraphics[width=0.32\hsize] {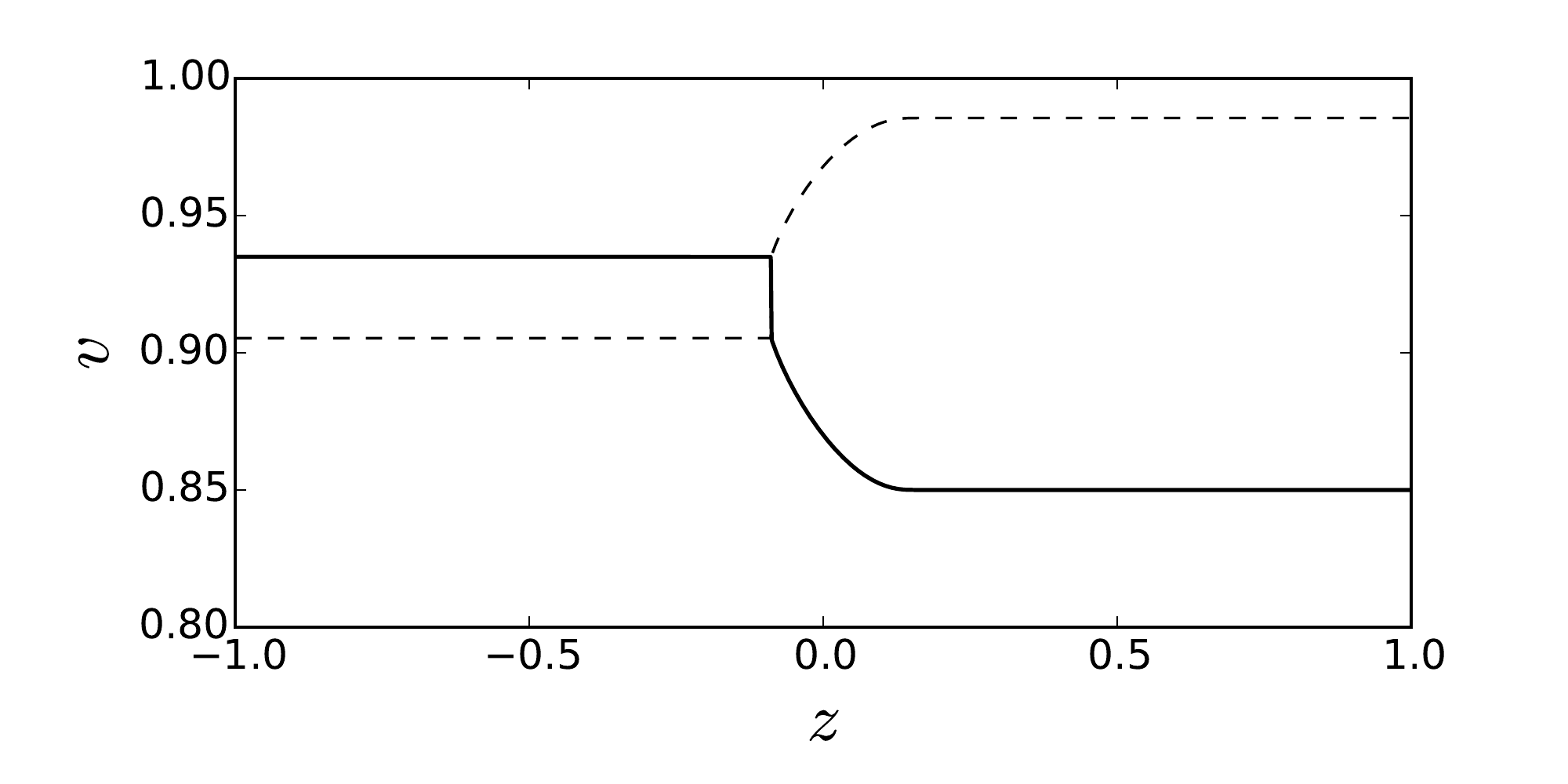}
		\includegraphics[width=0.32\hsize] {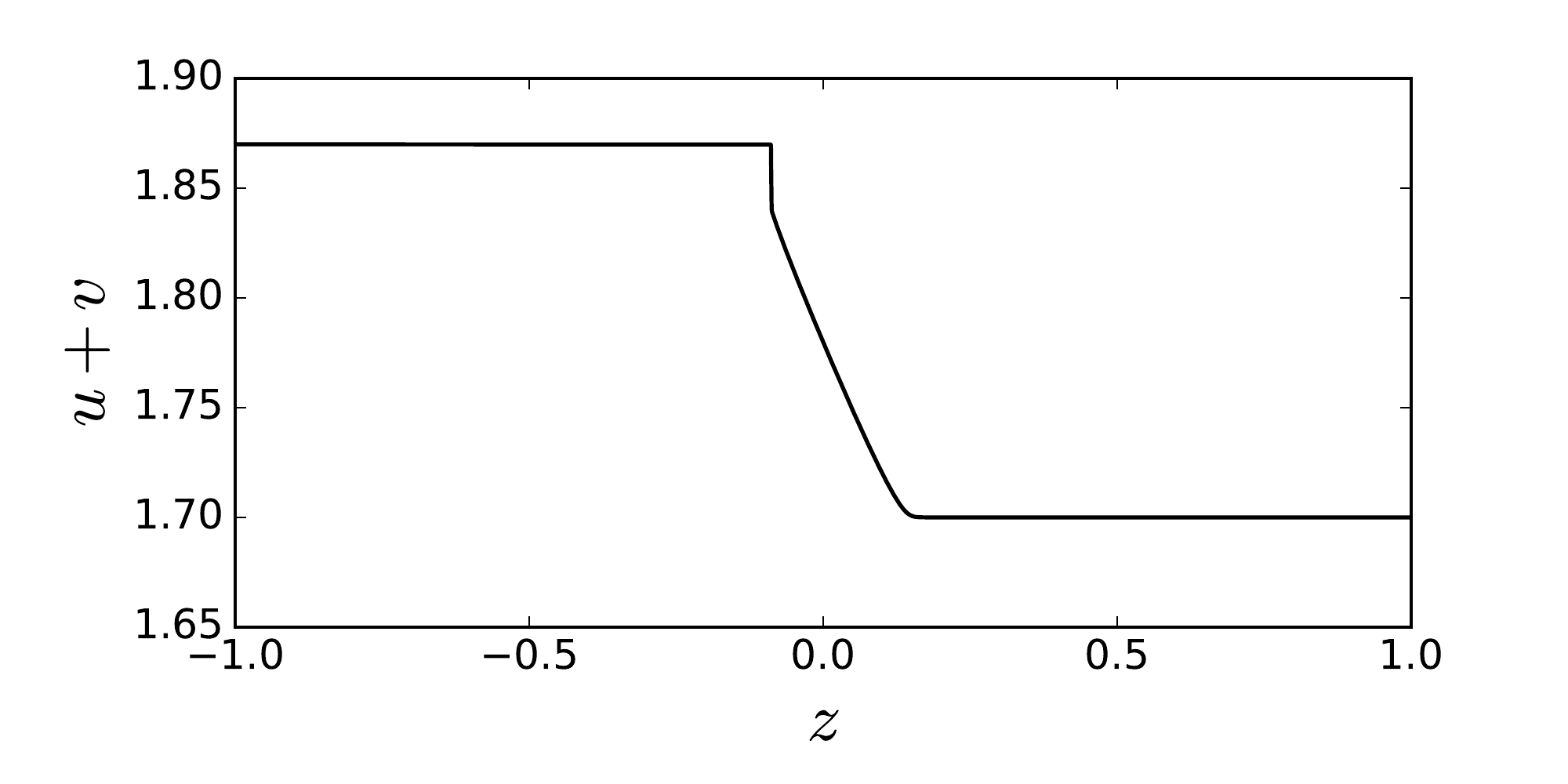}
		\includegraphics[width=0.32\hsize] {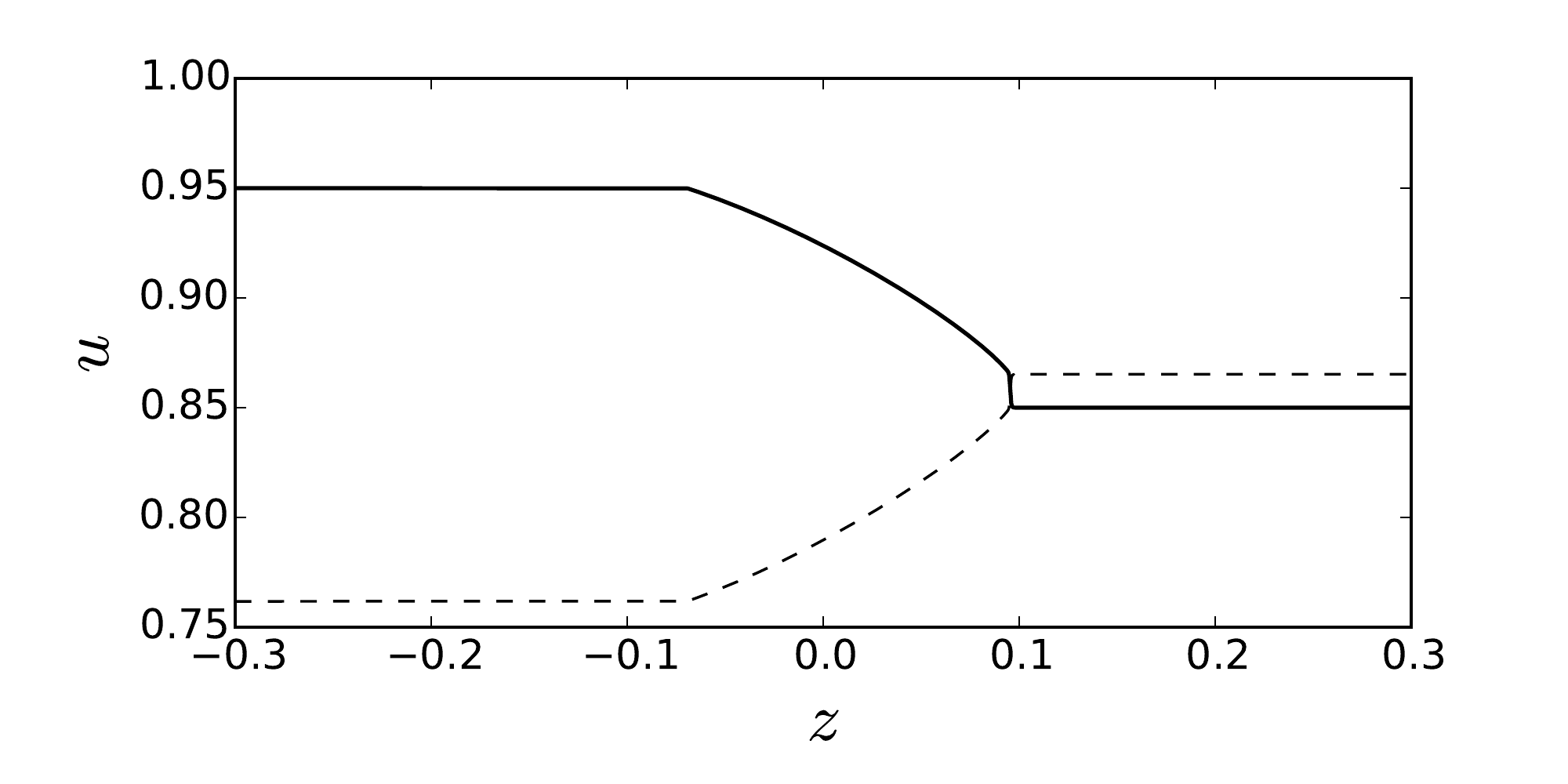}
		\includegraphics[width=0.32\hsize] {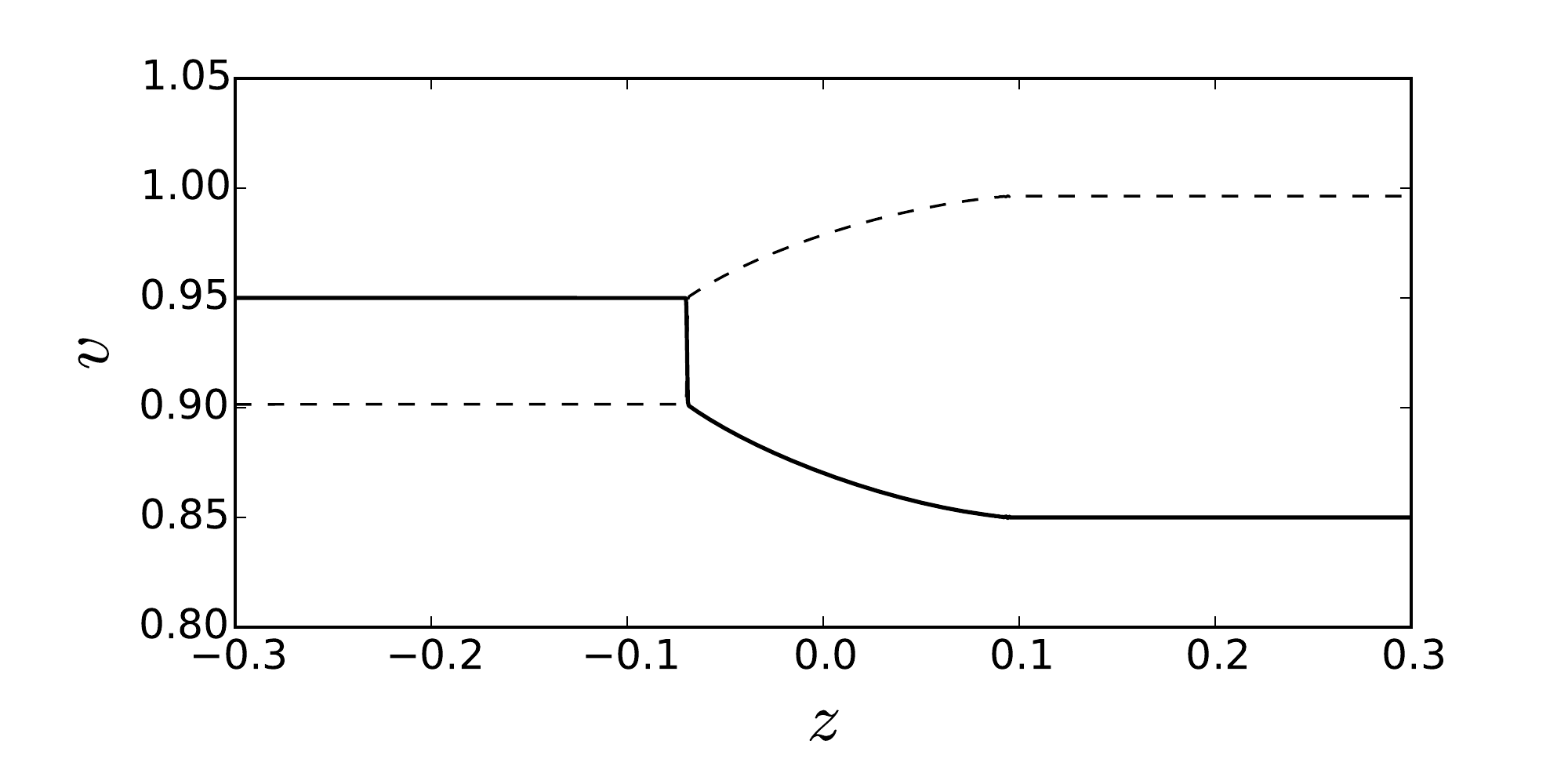}
		\includegraphics[width=0.32\hsize] {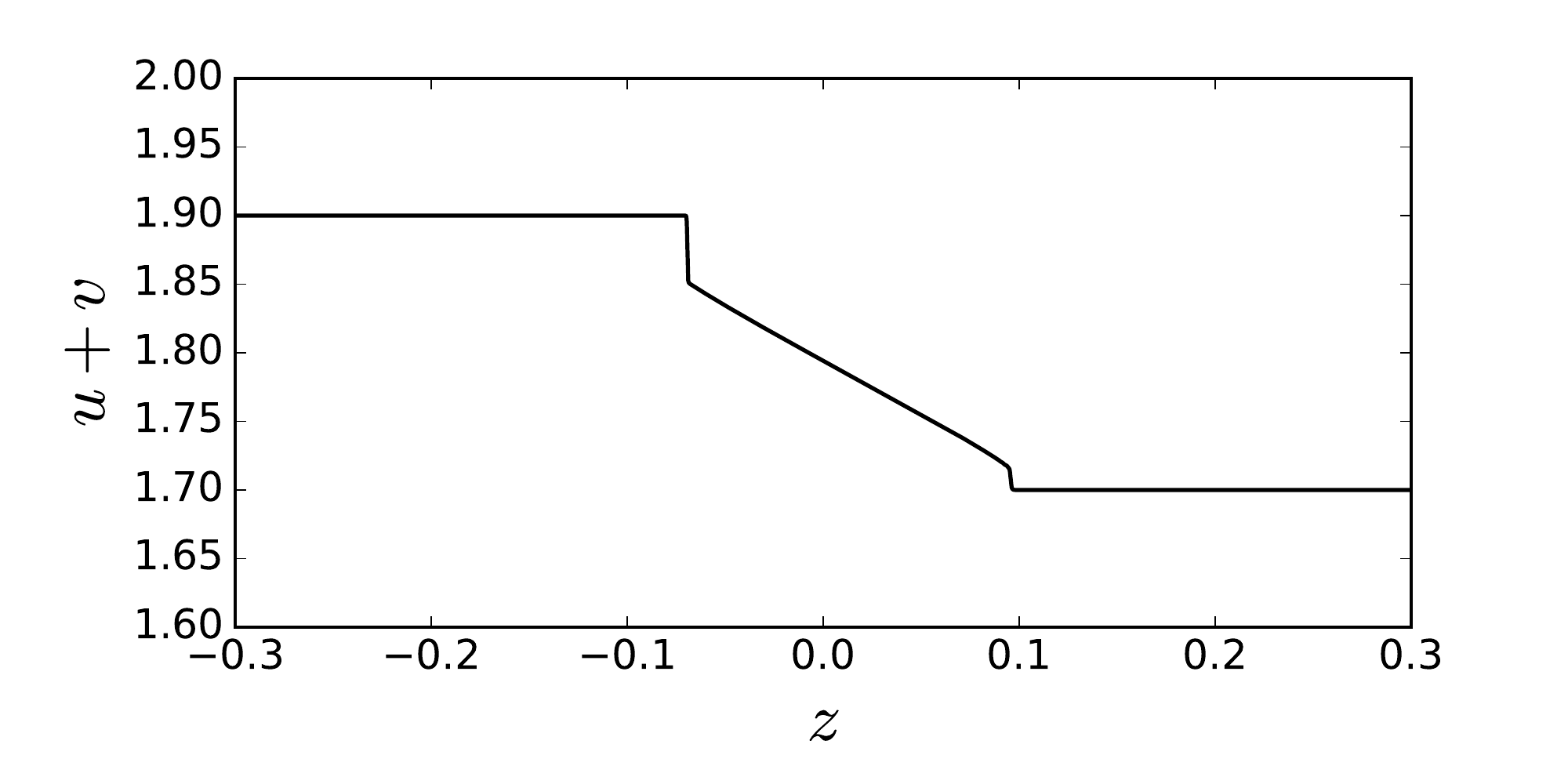}
		\caption{
			(Case B) Shock structure for $u_1 = 0.9$ (top), $u_1 = 0.935$ (middle) and $u_1 = 0.95$ (bottom). 
			The possible state just after the sub-shock (dotted curves) predicted by the RH conditions is also shown. 
			$u_0 = 0.85$ and $\tau = 1$. 
		}
		\label{fig:u_u0-085_u1-09}
	\end{center}
\end{figure*}
\subsection{Case C}
If we choose the state $\mathbf{U}_0 =(u_0, u_0)^T$ with  $0<u_0<0.536$, the relationship $\lambda^{(u)}_0 > \lambda^{(v)}_0$ holds. 
The characteristic velocity   $\lambda^{(v)}_1$ in the state $\mathbf{U}_1 = (u_1, u_1)$ coincides with the shock velocity at the critical characteristic velocity $s_*$, which is larger than the maximum characteristic velocity $s_*>\lambda^{(u)}_0$. 
We understand that there are two possibilities of the sub-shock formation  both for $s$ greater than $\lambda^{\max}_0$. 
The first  is the sub-shock appearing when $s > \lambda^{(u)}_0$. 
The second possibility is the sub-shock when $s > s_*$. 
The necessary condition \eqref{necessaria} is violated.
As a typical example, we show the shock velocity dependence of the characteristic velocities in the state $\mathbf{U}_1 = (u_1, u_1)^T$ for $u_0=0.3$ in Figure \ref{fig:toy_diagram_u0-03}. 
In the present case, $\mu_0=0.050$, $\lambda^{(v)}_0=0.0081$, $\lambda^{(u)}_0=0.09$ and $s_*=0.13$.

There is  another region of the state $\mathbf{U}_0 =(u_0, u_0)^T$ with  $1<u_0<1.06$, which belongs to the  Case C. 
The relationship $\lambda^{(v)}_0 > \lambda^{(u)}_0$ holds and the characteristic velocity   $\lambda^{(u)}_1$ in the perturbed state meets the shock velocity at the critical characteristic velocity $s_*$ larger than the maximum characteristic velocity $s_*>\lambda^{(v)}_0$.

\section{Numerical results on the shock wave structure}
In this section, we perform the numerical calculation on the shock structure in order to check the theoretical predictions of the sub-shock formation discussed in the previous section. 
We numerically solve the Riemann problem  with  the following initial  condition: 
\begin{equation*}
u (x, 0) = v (x, 0) 
= 
\begin{cases}
u_1 & (x < 0)\\
u_0 & (x \geq 0)\\
\end{cases}
\end{equation*}
\begin{figure*}[!]
	\begin{center}
		\includegraphics[width=0.32\hsize] {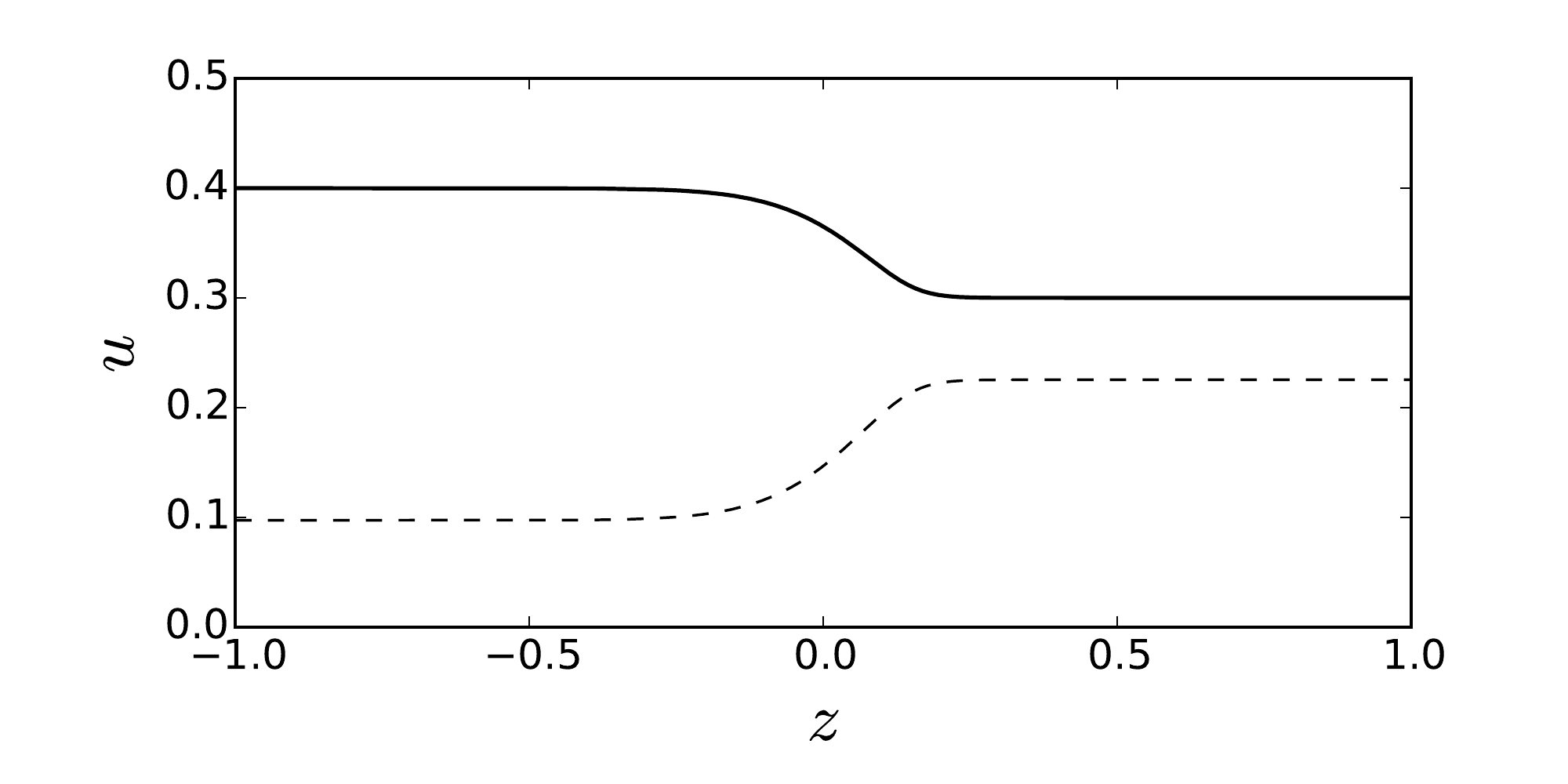}
		\includegraphics[width=0.32\hsize] {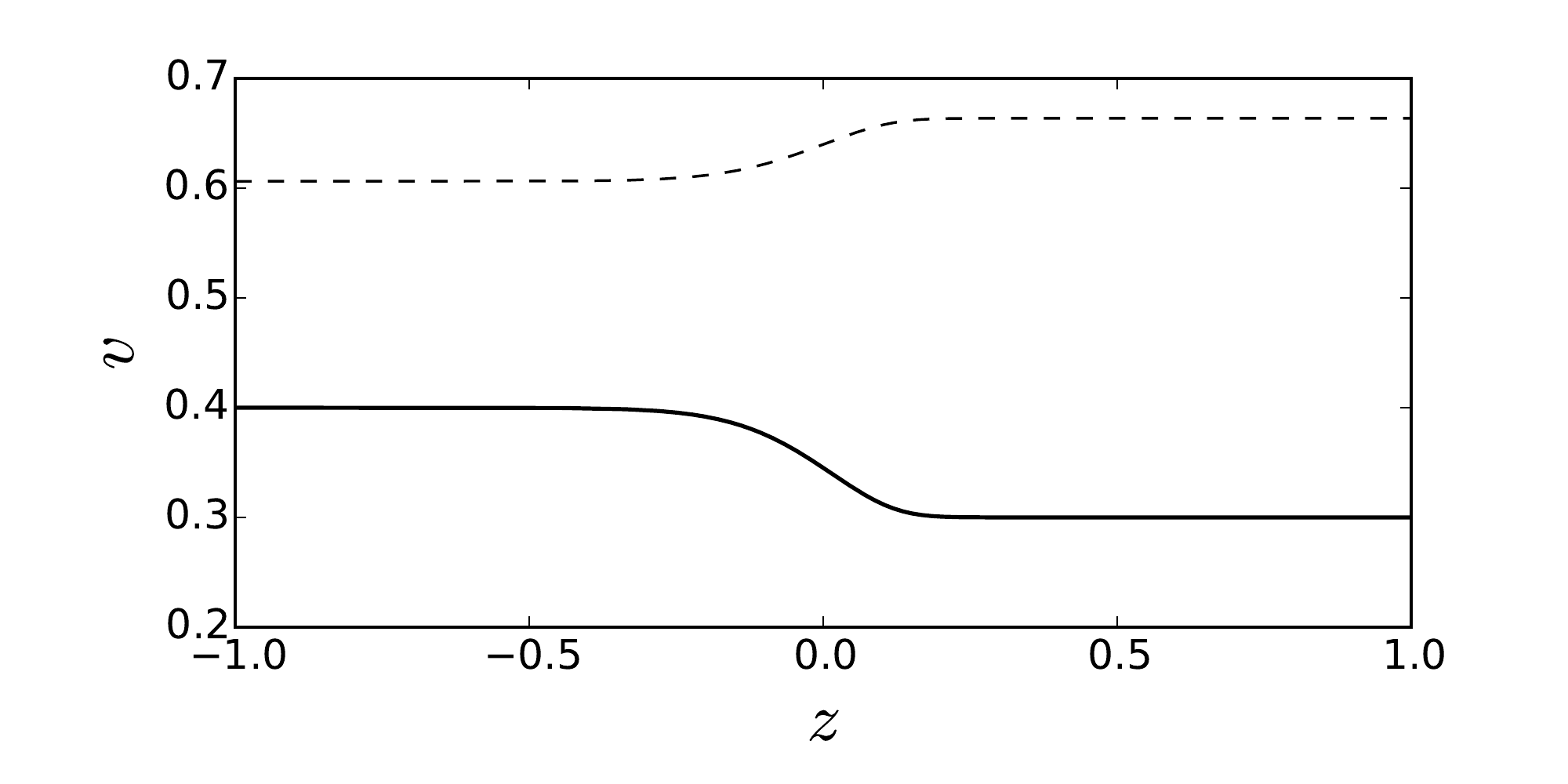}
		\includegraphics[width=0.32\hsize] {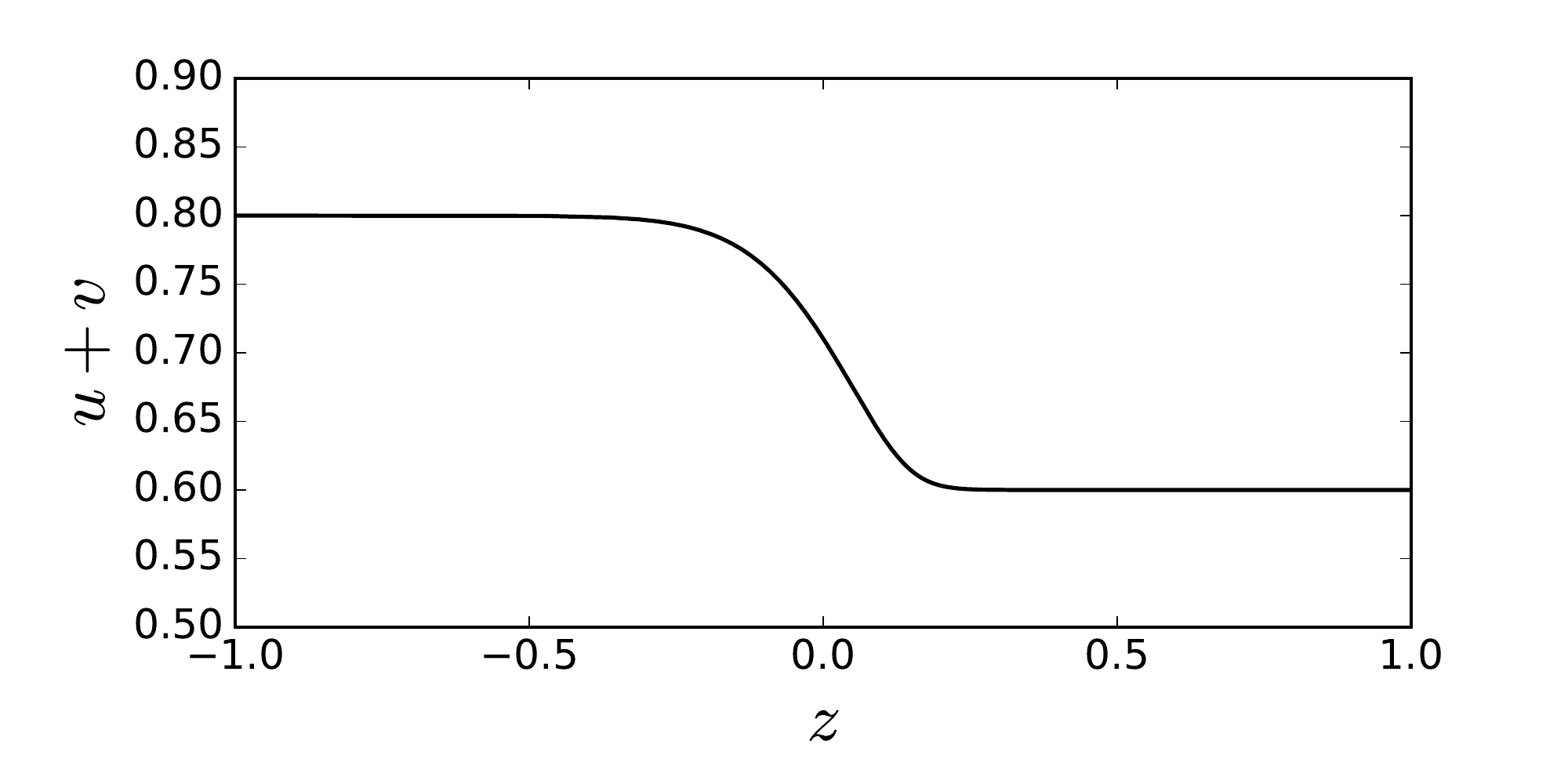}
		\includegraphics[width=0.32\hsize] {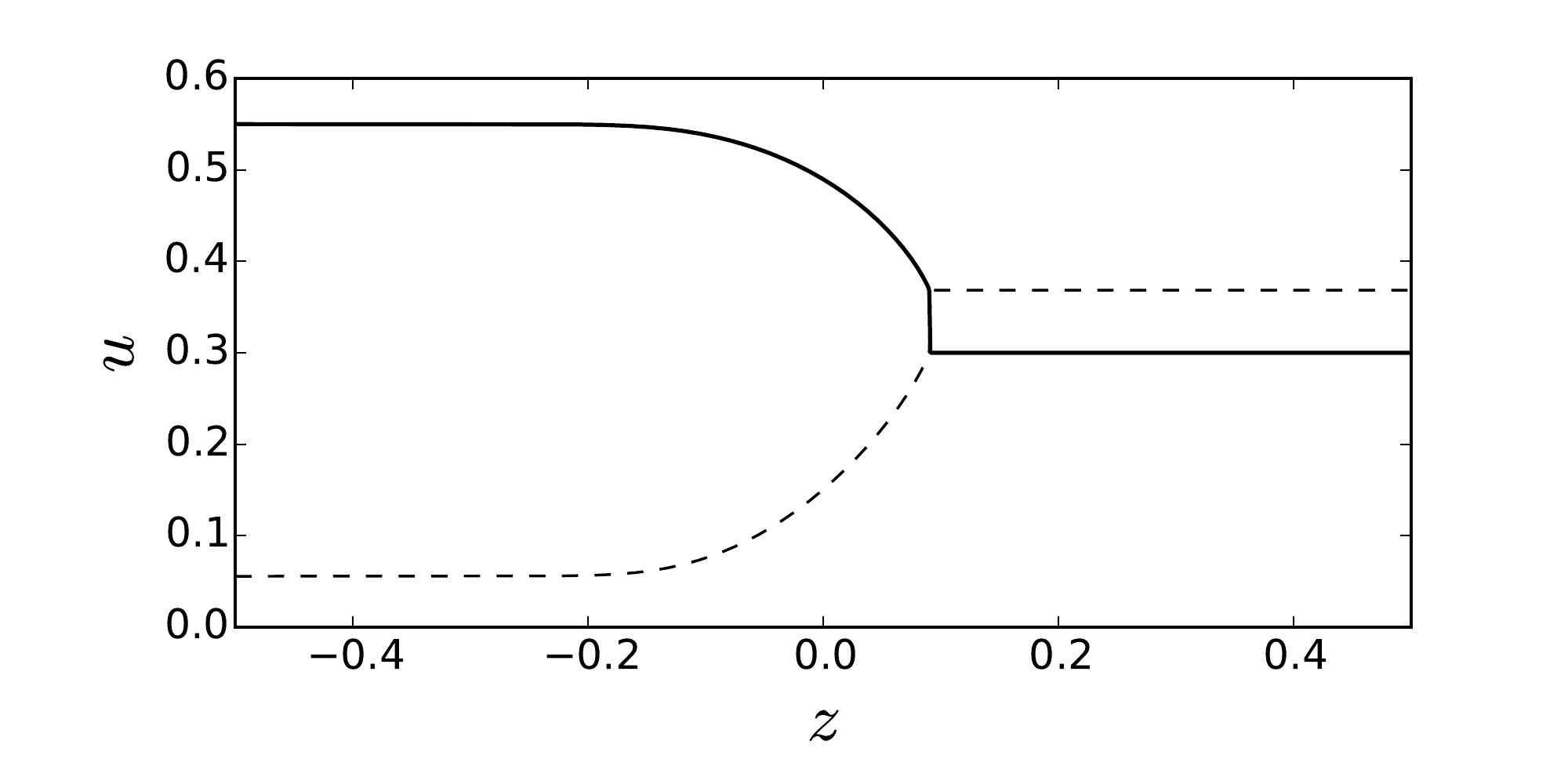}
		\includegraphics[width=0.32\hsize] {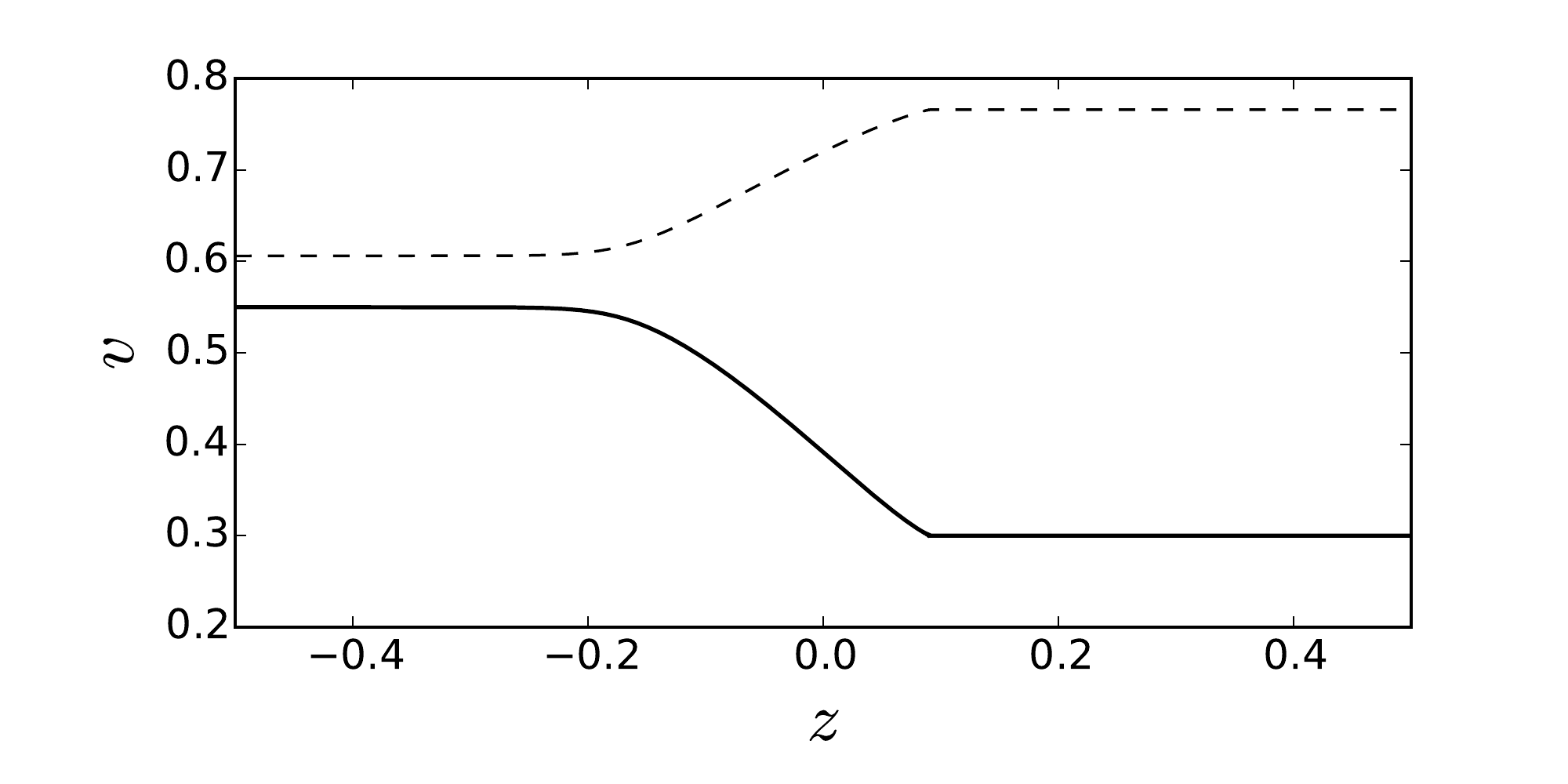}
		\includegraphics[width=0.32\hsize] {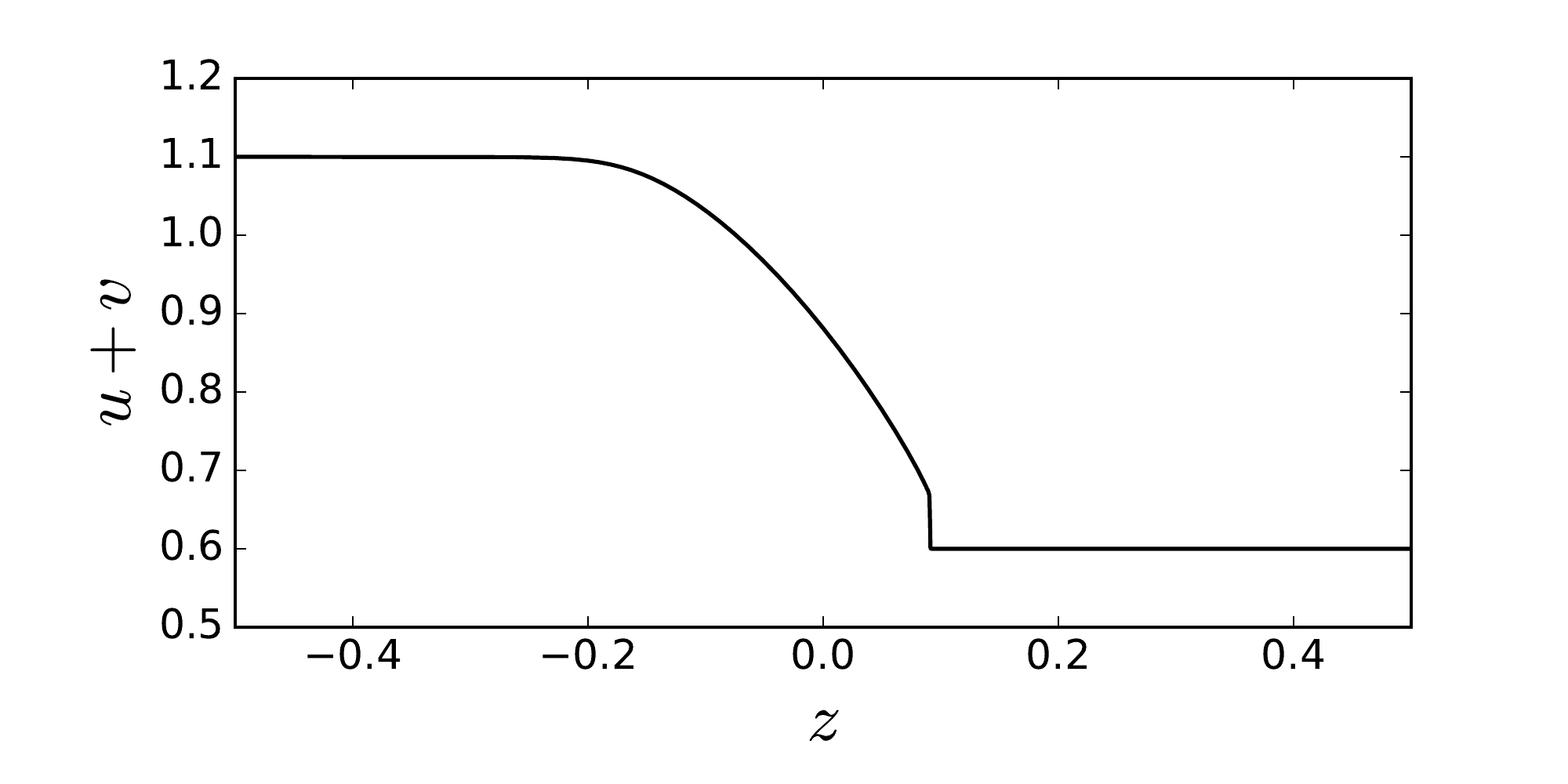}
		\includegraphics[width=0.32\hsize] {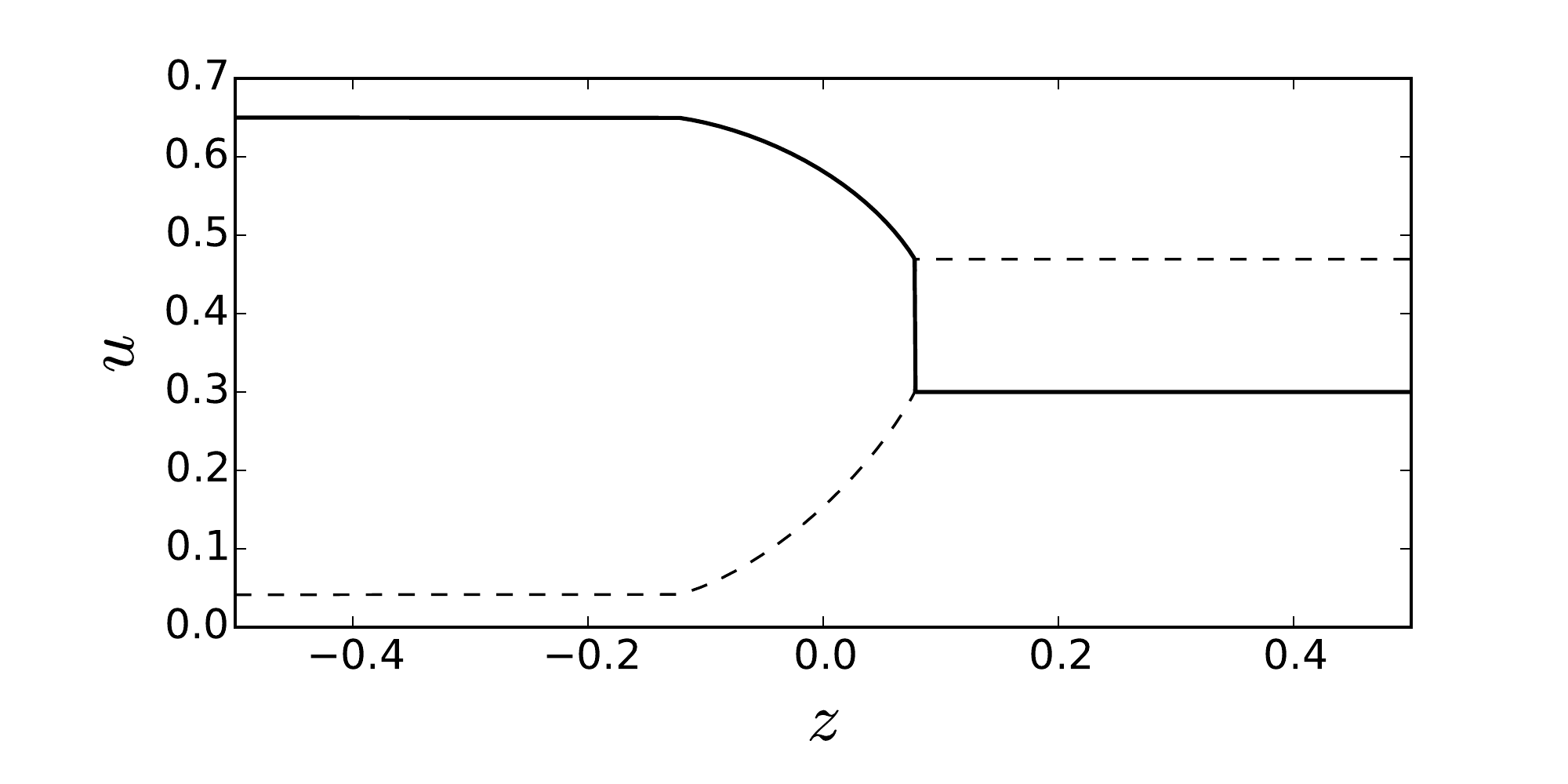}
		\includegraphics[width=0.32\hsize] {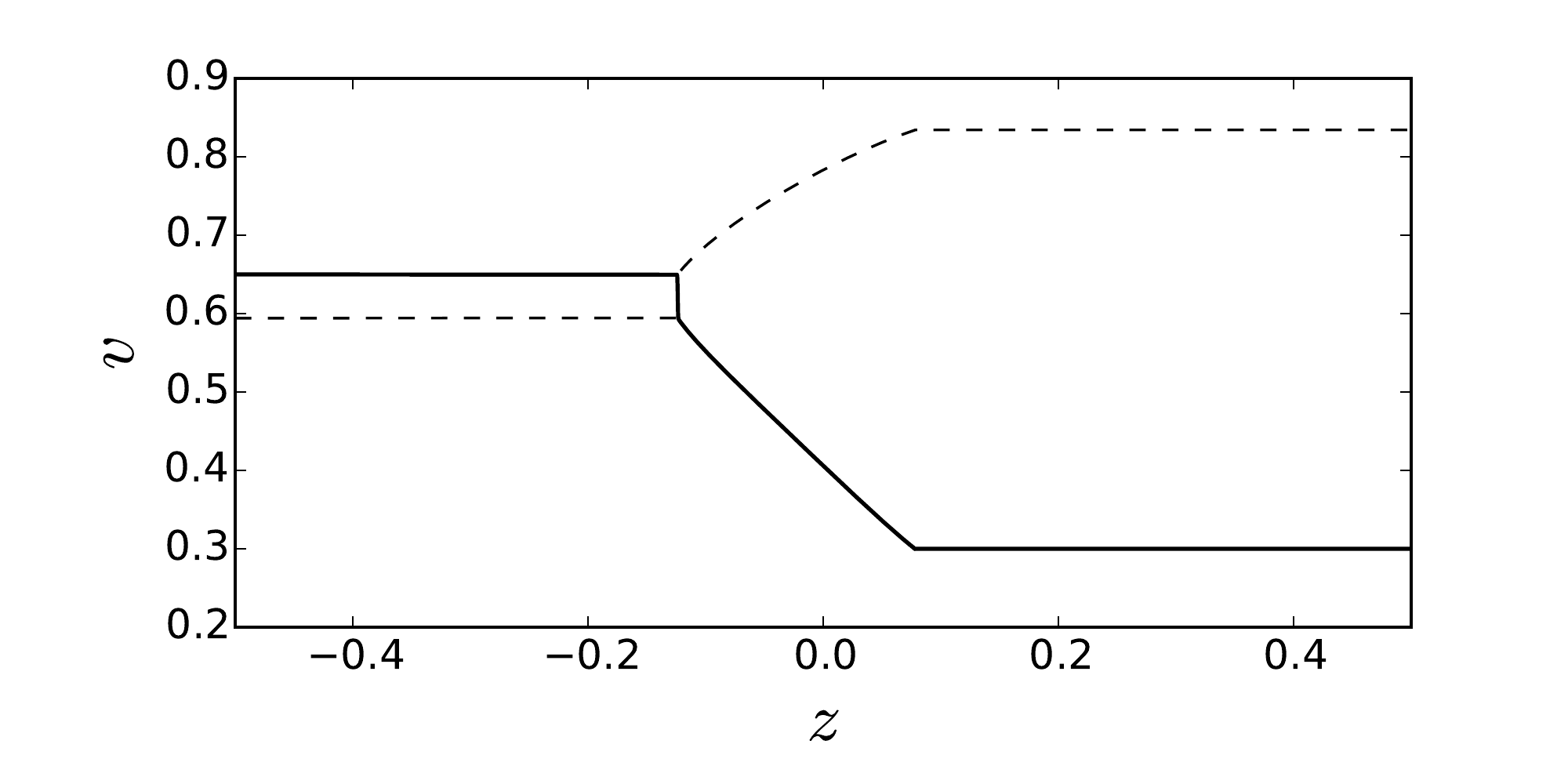}
		\includegraphics[width=0.32\hsize] {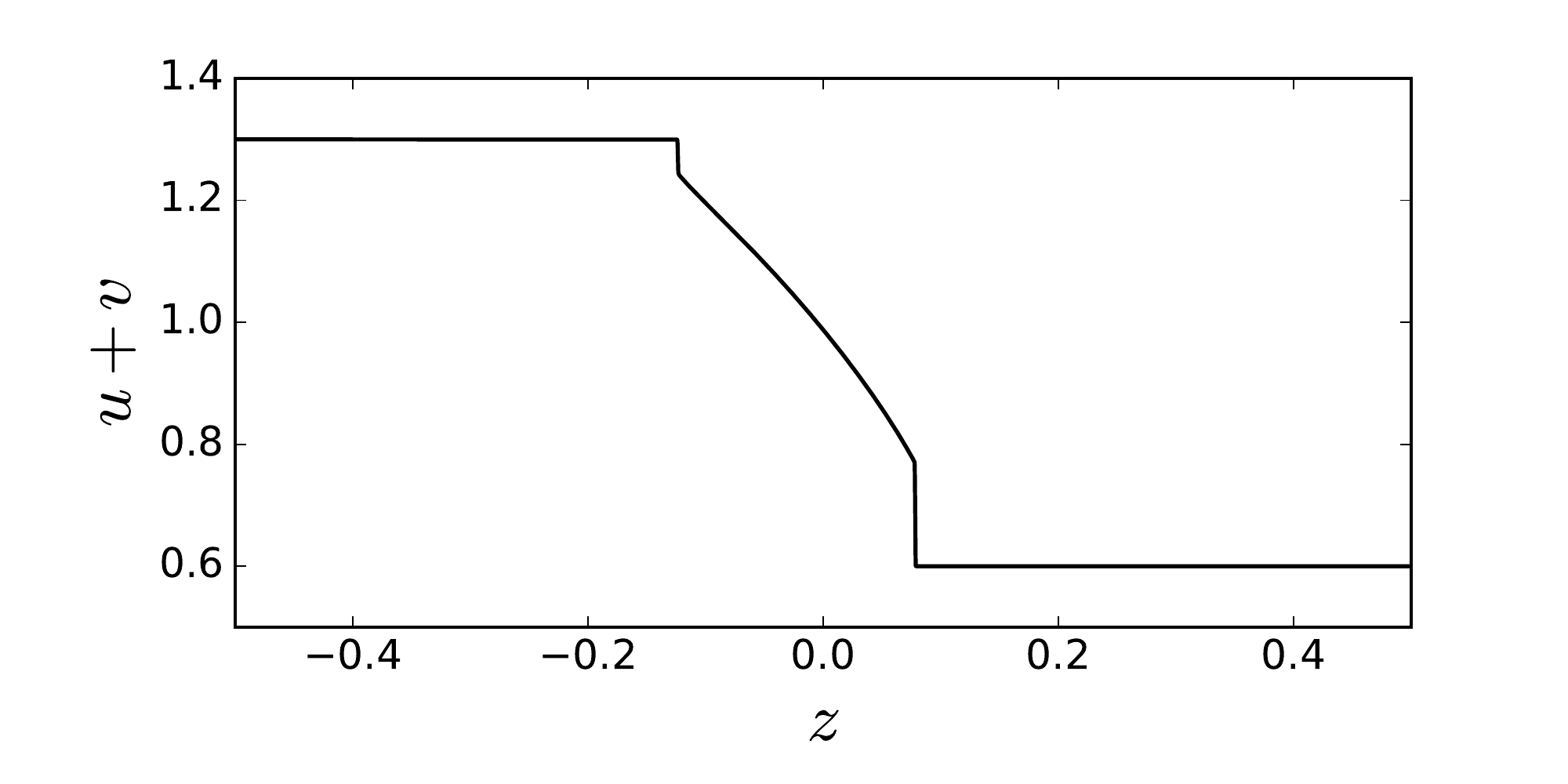}
		\caption{
			(Case C) Shock structure for $u_1 = 0.4$ (top), $u_1 = 0.55$ (middle) and $u_1 = 0.65$ (bottom). 
			The possible state just after the sub-shock (dotted curves) predicted by the RH conditions is also shown. 
			$u_0 = 0.3$ and $\tau = 1$.
		}
		\label{fig:u_u0-03_u1-04}
	\end{center}
\end{figure*}
with  $u_1 (u_0, s)$ satisfying RH conditions for the equilibrium subsystem \eqref{eq:toy-eqsub} 
and we analyze the shock-structure solution obtained 
%as the stationary profiles of $u$ and $v$ 
after long time according with the conjecture explained in Sec.~\ref{sec:conjecture}. 
Hereafter, we adopt $\tau=1$. 

As it is not easy to distinguish numerically a real sub-shock from a steep change of the profile, 
%to be sure that we have a sub-shock, 
we adopt a strategy used in a previous paper \cite{FMR}. 
This strategy is based on the fact that, if there exists a sub-shock, the two states $\mathbf{U}_-$ and $\mathbf{U}_+$  must satisfy the Rankine-Hugoniot for the full system, i.e.~\cite{MullerRuggeri,Dafermos}:
\begin{equation*}
-s[\![{\mathbf{U}}]\!] + [\![{\mathbf{F}({\mathbf{U}})}]\!] = 0,
\label{eq:general_RH}
\end{equation*}
where $[\![\psi
]\!]=\psi_{+}-\psi_{-}$ represents the jump of a generic quantity $\psi$ across
the (discontinuous) shock front. 
Here $\psi_{+}$ and $\psi_{-}$ are, respectively, the values of $\psi$ in the just right state and in the just left state of the jump. 
Therefore first we plot the profile of the shock structure and we consider  any point of the profile as the state just before a potential sub-shock $(u_{+},v_{+})^T$, and then, from the Rankine Hugoniot conditions for the full system \eqref{eq:field_toy2},
\begin{equation*}
\begin{split}
&s = \frac{u_{-}^2 + u_{+} u_{-} + u_{+}^2}{3}, \\
&s = \frac{v_{-}^4 + v_{-}^3 v_{+} + v_{-}^2 v_{+}^2 +  v_{-} v_{+}^3 + v_{+}^4}{5},  \\
\label{eq:RH_toy}
\end{split}
\end{equation*}
we associate  $(u_{+},v_{+})^T$ with a point $(u_{-},v_{-})^T$.  
%satisfying RH conditions \eqref{eq:RH_toy} with $(u_{+},v_{+})$. 
In this way we have two curves: the profile of the shock structure and the curve of potential state just after the sub-shock. 
If the two curves never meet, we understand that the profile of the shock structure is continuous and no sub-shock exists like in Figures \ref{fig:u_u0-115_u1-12}$_{1,2,4}$. 
If the two curve have two points in common like in Figure \ref{fig:u_u0-115_u1-12}$_5$, we understand that a sub-shock appears.

As a typical example of Case A, Figure \ref{fig:u_u0-115_u1-12} shows the numerical shock structure with or without a sub-shock for $u_1=1.2$ ($s=1.64$) and for $u_1=1.3$ ($s=1.89$).  
As was predicted, we have the continuous shock wave structure for $u_1 = 1.2$ and observe only one sub-shock for $u_1 = 1.3$.

As a typical example of Case B, Figure \ref{fig:u_u0-085_u1-09} shows the numerical shock structure for $u_1=0.9$ ($s=0.677$), for $u_1=0.935$ ($s=0.717$) and for $u_1=0.95$ ($s=0.735$). 
We see the continuous shock structure for $u_1 = 0.9$. 
It should be emphasized that we observe the sub-shock formation for $u_1 = 0.935$ which satisfies the RH conditions for the sub-shock and that this is clearly a counter example of the sub-shock slower than the maximum unperturbed characteristic velocity. 
We see also the multiple sub-shock for $u$ and $v$ for $u_1 = 0.95$. 

As a typical example of Case C, Figure \ref{fig:u_u0-03_u1-04} shows the numerical shock structure for $u_1=0.4$ ($s=0.069$), for $u_1=0.55$ ($s=0.11$) and for $u_1=0.65$ ($s=0.15$). 
As was predicted, we see the continuous shock wave structure for $u_1 = 0.4$, the structure with one sub-shock for $u$ for $u_1 = 0.55$ and the formation of the multiple sub-shock for $u$ and $v$ for $u_1 = 0.65$.

\section{Summary and concluding remarks}

In this paper,
first, we have shown that ET for a rarefied polyatomic gas with 14 independent variables does not predict the sub-shock formation with slower shock velocity than the maximum unperturbed characteristic velocity. 
Second, we have shown an example of the clear sub-shock formation with slower shock velocity than the maximum characteristic velocity by adopting a simple $2 \times 2$ hyperbolic dissipative system that satisfies all requirements of the ET theory. 
We have concluded that the requirements of   the entropy principle, the convexity of the entropy and the Shizuta-Kawashima condition, are not enough to characterize the property on the sub-shock formation of ET. 

Therefore, if we  conjecture that ET theories have this strange beautiful property such that the sub-shock appears only for the shock velocity greater than the maximum characteristic velocity, there must  exist some special property of differential system of  ET theories, which is still obscure.

If we multiply the system \eqref{struttura} by the left eigenvector $\mathbf{l}$ of $\mathbf{A}$ corresponding to a given eigenvalue $\lambda$, we obtain
\begin{equation*}
\mathbf{l} \cdot \frac{d \mathbf{U}}{d z} = \frac{\mathbf{l}\cdot \mathbf{f}}{\lambda -s}. 
\end{equation*}
To make the solution regular, when the eigenvalue $\lambda$ approaches to $s$, $\mathbf{l}\cdot \mathbf{f}$ also must tend to 0.
This means that the differential system of ET theories needs to satisfy some special condition between productions and the main part of the operator and this condition may be  more restrictive than the K-condition. 
The identification of this condition is still an open problem and we will try to give an answer in the future.

\section*{Acknowledgments}
\small{This work was partially supported by JSPS KAKENHI Grant Number JP16K17555 (S. T.) and by National Group of Mathematical Physics GNFM-INdAM (T. R.).}


\begin{thebibliography}{99}
\bibitem{sign_entropy}
We adopt different definition of the sign of the entropy from the one adopted in~\cite{MentrelliRuggeri}.
\end{thebibliography}
\end{document}